\pdfoutput=1

\documentclass[11pt]{article}

\usepackage[]{acl}

\usepackage{times}
\usepackage{latexsym}
\usepackage{amsmath}
\usepackage{graphicx}
\usepackage{amssymb} 
\usepackage{svg}
\usepackage{multirow} 
\usepackage{hyperref}

\usepackage[T1]{fontenc}

\usepackage[utf8]{inputenc}

\usepackage{microtype}

\usepackage{inconsolata}
\usepackage{graphicx}

\usepackage[table]{xcolor}
\definecolor{lightgray}{gray}{0.9}

\title{DisastIR: A Comprehensive Information Retrieval Benchmark for\\Disaster Management}

\author{
\textbf{Kai Yin}\textsuperscript{1} \quad 
\textbf{Xiangjue Dong}\textsuperscript{1} \thanks{Corresponding author.} \quad 
\textbf{Chengkai Liu}\textsuperscript{1} \footnotemark[1] \quad 
\textbf{Lipai Huang}\textsuperscript{1} \\
\textbf{Yiming Xiao}\textsuperscript{1} \quad 
\textbf{Zhewei Liu}\textsuperscript{2} \quad 
\textbf{Ali Mostafavi}\textsuperscript{1} \quad 
\textbf{James Caverlee}\textsuperscript{1} \\
\textsuperscript{1}Texas A\&M University \quad 
\textsuperscript{2}University of Toronto \\
\texttt{\{kai\_yin, xj.dong, liuchengkai, lipai.huang, yxiao, mostafavi, caverlee\}@tamu.edu} \\
\texttt{zwei.liu@utoronto.ca} \\
}

\begin{document}
\maketitle
\begin{abstract}
Effective disaster management requires timely access to accurate and contextually relevant information. Existing Information Retrieval (IR) benchmarks, however, focus primarily on general or specialized domains, such as medicine or finance, neglecting the unique linguistic complexity and diverse information needs encountered in disaster management scenarios. To bridge this gap, we introduce \textbf{DisastIR}, the first comprehensive IR evaluation benchmark specifically tailored for disaster management. DisastIR comprises 9,600 diverse user queries and more than 1.3 million labeled query-passage pairs, covering 48 distinct retrieval tasks derived from six search intents and eight general disaster categories that include 301 specific event types. Our evaluations of 30 state-of-the-art retrieval models demonstrate significant performance variances across tasks, with no single model excelling universally. Furthermore, comparative analyses reveal significant performance gaps between general-domain and disaster management-specific tasks, highlighting the necessity of disaster management-specific benchmarks for guiding IR model selection to support effective decision-making in disaster management scenarios. All source codes and DisastIR are available at \href{https://github.com/KaiYin97/Disaster_IR}{this repository}. 

\end{abstract}

\section{Introduction}
\label{sec: intro}

Natural disasters and technological crises cause severe threats to human lives, infrastructure, and the environment, necessitating timely and effective management responses \citep{dong2020integrated, yin2023integrated, liu2024artificial}. In such critical scenarios, stakeholders, including emergency responders, government agencies, and the general public, require rapid access to reliable and contextually relevant information to make informed decisions \citep{jayawardene2021role, abbas2025exploring}. Information Retrieval (IR) systems thus play a critical role in disaster management, where rapid, accurate access to relevant information can significantly impact emergency response outcomes and decision-making efficacy \citep{dirsm2020, taqe2023, search2024}.

\begin{figure}[tbp]
    \centering
     \includegraphics[width=\linewidth]{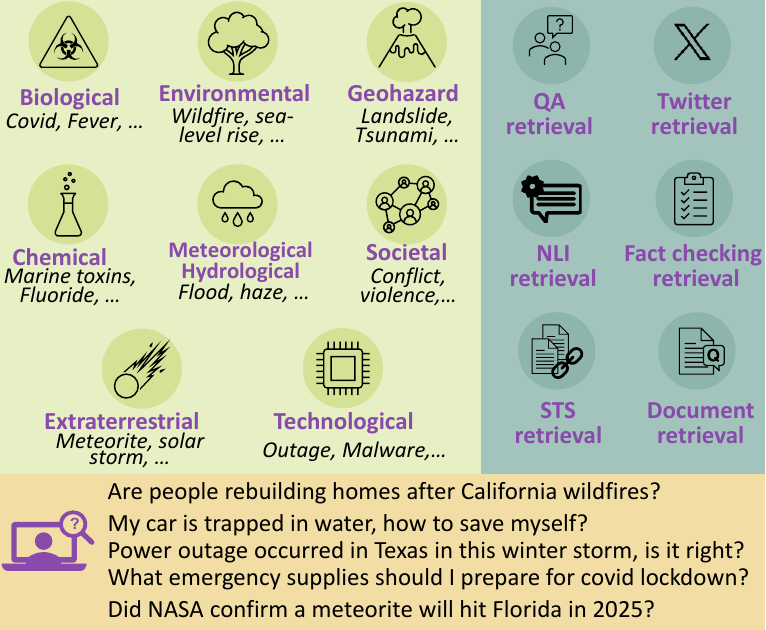}
    \caption{Examples of user queries across diverse search intents and event types during disaster management.}
    \label{Fig: real_world_query}
\vspace{-11pt}
\end{figure}

\begin{figure*}[ht]
    \centering
    \includegraphics[width=\textwidth]{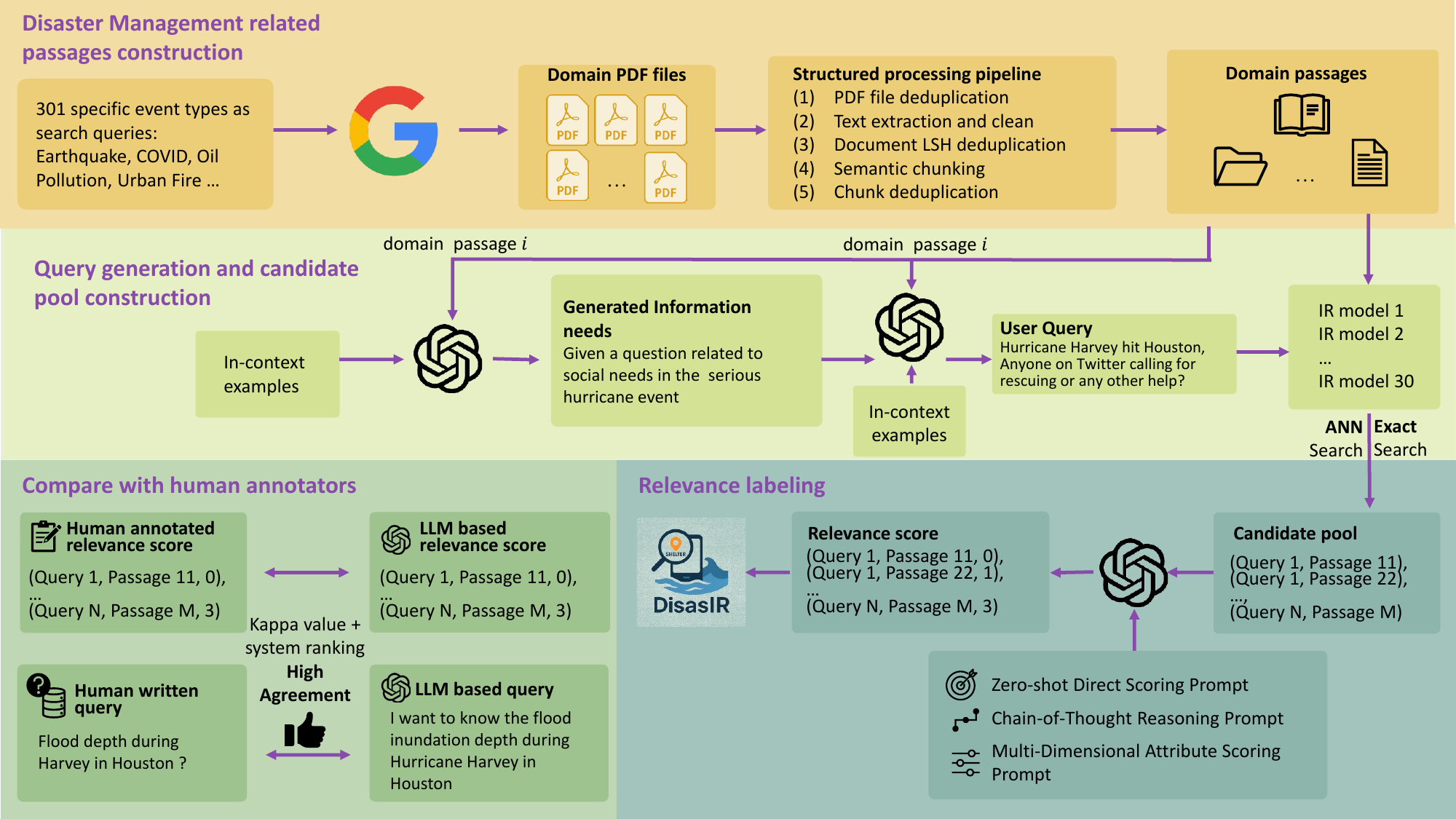}
    \caption{Proposed framework to develop DisastIR from scratch.}
    \label{fig:workflow}
\vspace{-10pt}
\end{figure*}

Information needs during real-world disasters are highly diverse (Figure \ref{Fig: real_world_query}), including intents such as question answering, rumor verification, social media monitoring, and evidence retrieval \citep{purohit2014emergency, imran2015processing, zubiaga2018detection}. These varied intents require tailored retrieval behavior \citep{asai2022task, su2022one, lee2024gecko} and understanding of ``relevance'' \citep{dai2022promptagator}. In addition, different types of disasters (Figure \ref{Fig: real_world_query}), such as geohazards, biological threats, and technological failures, differ significantly in terminology, phrasing, and discourse styles \citep{andharia2020disaster, undrr2020hazard, bromhead2021disaster}. This complexity presents significant challenges for retrieval systems aiming to serve real-world disaster response scenarios.

However, existing retrieval benchmarks primarily target general-domain tasks, such as BEIR \citep{thakur2021beir}, or focus on specific domains like medicine \citep{mirage2024} and finance \citep{tang2024finmteb}. They are not designed to reflect the search task diversity and domain-specific demands of disaster management scenarios. As a result, current IR evaluation benchmarks offer limited guidance for selecting retrieval models in disaster management applications.

To address this gap, we present DisastIR, the first comprehensive IR benchmark tailored to disaster management. DisastIR evaluates retrieval models across 48 distinct tasks, defined by combinations of six real-world search intents and eight general disaster event types, covering a total of 301 specific event types (see Section~\ref{sec:task}).

DisastIR is built on a systematically constructed disaster management-specific corpus, developed through extensive web crawling, semantic chunking, and deduplication (Section~\ref{sect: domain_knowledge_chunk}). To simulate realistic information needs, we use a large language model (LLM)\footnote{The LLM used in this work is GPT-4o-mini.} to generate diverse, contextually grounded user queries (Section~\ref{sec: user_query_gen}). Candidate passages are aggregated from multiple state-of-the-art (SOTA) retrieval models (Section~\ref{sec: candidate_pool}), and query-passage pairs are annotated using LLMs with three different designed prompts whose outputs are ensembled for robust relevance labeling (Section~\ref{sec: reve_label}).

To ensure annotation quality and evaluation reliability, we validate LLM-generated relevance labels against human annotations, observing substantial agreement (average Cohen’s kappa = 0.77; see Section~\ref{sec: llm_human_label}). We also compare LLM-generated and human-written queries across all 48 tasks (Section~\ref{sec:llm_human_query}) and find highly consistent evaluation results (Kendall’s $\tau$ = 0.93), supporting the use of synthetic queries and relevance labels in DisastIR.

Using DisastIR, we benchmark 30 open-source retrieval models of varying sizes, architectures, and backbones under both exact and approximate nearest neighbor (ANN) search settings (Section~\ref{sec:experiment_steup}). Our results show that no single model consistently outperforms others across all disaster management-related retrieval tasks (Section~\ref{section: perform_48_tasks}). We also observe substantial performance gaps between general-domain benchmarks (e.g., MTEB \citep{muennighoff2022mteb}) and DisastIR (Section~\ref{sec:compare_with_mteb}), highlighting the need for a domain-specific benchmark to guide reliable and effective retrieval model selection in disaster management scenarios.

The contributions of this work are as follows:
\begin{itemize}
\vspace{-4pt}
\item[(1)] We release DisastIR, the first IR benchmark tailored to disaster management. It includes a systematically constructed evaluation corpus of 239,704 passages and 9,600 user queries, with over 1.3 million annotated query-passage pairs across 48 retrieval tasks spanning diverse search intents and disaster event types.
\vspace{-4pt}
\item[(2)] We conduct a comprehensive evaluation of 30 open-source retrieval models under both exact and ANN search settings, offering practical guidance for model selection based on task requirements and computational constraints in disaster management scenarios.
\vspace{-4pt}
\item[(3)] We empirically demonstrate substantial performance gaps between general-domain and disaster management-specific retrieval, underscoring the necessity of disaster management-specific IR evaluation benchmarks.
\end{itemize}

\section{Related work}
Existing IR benchmarks target mainly general-purpose or specialized domains, such as medicine and finance. BEIR \citep{thakur2021beir} evaluates zero-shot retrieval models across 18 tasks, such as fact verification, QA, and scientific document ranking. Instruction-based benchmarks like FollowIR \citep{weller2024followir}, InstructIR \citep{instructir2023}, and MAIR \citep{zhang2024mair} reformulate IR tasks using natural language instructions. Some domain-specific IR benchmarks, such as MIRAGE \citep{mirage2024} and FinMTEB \citep{tang2024finmteb}, focus on biomedical and financial domains. While effective in their respective domains, they fail to capture the linguistic and contextual patterns in disaster management areas. 

Despite the critical role of information retrieval in disaster management, existing benchmarks are limited in scope, scale, and task diversity. Prior datasets—such as the FIRE IRMiDis track \citep{basu2017overview} and event-specific corpora from disasters in Nepal, Italy, and Indonesia \citep{khosla2017microblog, basu2019extracting, taqe2023}—primarily focus on Twitter microblogs, targeting short-text retrieval or keyword matching with narrow task coverage. Case-based systems like \citet{search2024} use proprietary data for concept-based retrieval in search and rescue planning. These benchmarks typically rely on single-source or scenario-specific data and lack support for realistic, multi-intent retrieval. In contrast, DisastIR provides a large-scale, multi-intent, and multi-source benchmark covering diverse disaster types and information needs, enabling comprehensive evaluation in real-world contexts.

\section{DisastIR: Disaster Management Information Retrieval Benchmark}

\subsection{Overview}

The construction of DisastIR follows a four-stage pipeline, as illustrated in Figure~\ref{fig:workflow}: (1) disaster management corpus construction, (2) user query generation, (3) candidate pool development, and (4) relevance labeling. DisastIR is built upon a large-scale, high-quality corpus of disaster management-related passages covering diverse event types. User queries are generated by prompting an LLM with these domain passages as context, targeting different search intents. Relevance scores for each query-passage pair are then assigned by the LLM.

\subsection{Evaluation Task}
\label{sec:task}

To evaluate how well retrieval models address diverse user intents and disaster contexts, DisastIR defines six search intents and eight general disaster event types, resulting in 48 distinct retrieval tasks. 
 
Specifically, 301 specific event types are identified spanning eight general categories: Biological (Bio), Chemical (Chem), Environmental (Env), Extraterrestrial (Extra), Geohazard (Geo), Meteorological \& Hydrological (MH), Societal (Soc), and Technological (Tech) \citep{undrr2020hazard}. See Figure \ref{Fig: real_world_query} for examples of specific event types belonging to each general disaster event type. 
 
Six distinct search intents are included, inspired by prior benchmarks such as BEIR \citep{thakur2021beir}, BERRI \citep{asai2022task}, MEDI \citep{su2022one}, and MAIR \citep{sun2024mair}: question-answer (QA) retrieval, Twitter retrieval, Fact Checking (FC) retrieval, Natural Language Inference (NLI) retrieval, and Semantic Textual Similarity (STS) Retrieval. For QA, we further distinguish between retrieving relevant passages (QA) and retrieving relevant documents (QAdoc), following common practice in prior work \citep{kwiatkowski2019natural, khashabi2021gooaq, xu2024bmretriever}.\footnote{A passage refers to a single chunk with limited token length, while a document denotes a full source file, which may be segmented into multiple passages.} Due to token limitations in many retrieval models -- especially encoder-based ones -- it is often infeasible to encode full documents directly. To address this, we prompt an LLM to summarize each document and include the summary in the corpus as a proxy for the original document.

\subsection{Domain knowledge corpus construction}
\label{sect: domain_knowledge_chunk}

To construct the domain knowledge corpus, we perform a large-scale web crawling using 301 disaster event types as search queries, collecting domain-specific PDF documents from publicly available sources. A structured pipeline is then applied to convert raw PDFs into clean, retrieval-ready passages: (1) exact-URL deduplication, (2) text extraction and preprocessing, (3) document-level near-duplicate removal using locality-sensitive hashing (LSH), (4) semantic chunking, and (5) embedding-based near-duplicate filtering. The full pipeline is described in Appendix~\ref{appendix:pdf_pipeline}.

\subsection{User Query Generation} 
\label{sec: user_query_gen}

A key challenge in constructing domain-specific IR evaluation datasets is generating user queries that reflect real information needs \citep{rahmani2024synthetic}. With the advent of LLMs, it is now feasible to synthesize high-quality, diverse, and contextually grounded queries by prompting models with domain-specific passages \citep{alaofi2023can, rajapakse2023improving, rahmani2024synthetic}.

In this work, we propose a two-stage few-shot prompting strategy to generate user queries based on disaster management passages. In the first stage, an LLM is prompted to brainstorm diverse information need statements grounded in the content of the given passage. In the second stage, given a randomly selected information need and the associated passage, the LLM generates a user query and a directly relevant passage as shown below:
\vspace{-5pt}

\[
\begin{aligned}
&LLM_{query}\bigl(
   \underbrace{LLM_{info}(P_{IN},P_{seed})}_{\displaystyle \text{information need}},
   \;P_{QG},\;P_{seed}
  \bigr) \\
&~~~~~~~~~~~\quad\longrightarrow\;(q,\;psg) ~~~~~~~~~~~~~~~~~~~~~~~~~~~~~~~~~(1)
\end{aligned}
\]
where \( LLM_{info} \) and \( LLM_{query} \) are LLMs prompted to generate retrieval information needs statements and the query-passage pair respectively, \( P_{IN} \) and \( P_{QG} \) are prompts for information needs and query generation, \( P_{seed} \) is the domain passage, \( q \) is the synthesized user query, and \( psg \) is the corresponding relevant passage. 

To ensure generated queries align with the core characteristics and objectives of each search intent, we design intent-specific prompts for both stages of query generation. The full prompt templates for each intent are provided in Appendix~\ref{appendix:prompts_query_gen}.

For each search task, we generate 200 unique user queries by prompting an LLM with randomly sampled domain-specific passages, resulting in 9,600 queries. 
The final corpus combines disaster management-related passages from Section~\ref{sect: domain_knowledge_chunk} with generated passages to reflect various search intents. Some tasks, such as Twitter, NLI, and FC retrieval, require passage types with distinct styles and semantics. Including generated passages ensures the corpus can support realistic evaluation across diverse retrieval scenarios.\footnote{Relevance scores of query-generated passage pairs are also evaluated instead of directly giving them the highest relevance score.}

\begin{table*}[ht]
\centering
\small
\renewcommand{\arraystretch}{1.15}
\setlength{\tabcolsep}{3pt}
\resizebox{\textwidth}{!}{%
\begin{tabular*}{\textwidth}{@{\extracolsep{\fill}}@{\hspace{5pt}}l|cccccc@{\hspace{5pt}}}
\hline
\textbf{}
  & \textbf{QA}
  & \textbf{QAdoc}
  & \textbf{Twitter}
  & \textbf{FC}
  & \textbf{NLI}
  & \textbf{STS} \\
\hline
\textbf{Bio}   & 26651 (133.3) & 25335 (126.7) & 35182 (175.9) & 23987 (119.9) & 25896 (129.5) & 27065 (135.3) \\
\textbf{Chem}  & 26885 (134.4) & 26032 (130.2) & 34186 (170.9) & 24592 (123.0) & 27856 (139.3) & 26787 (133.9) \\
\textbf{Env}   & 26685 (133.4) & 25930 (129.7) & 33243 (166.2) & 25805 (129.0) & 25207 (126.0) & 27048 (135.2) \\
\textbf{Extra}& 26807 (134.0) & 25598 (128.0) & 33202 (166.0) & 24363 (121.8) & 26399 (132.0) & 27313 (136.6) \\
\textbf{Geo}   & 27140 (135.7) & 26573 (132.9) & 35503 (177.5) & 27864 (139.3) & 28210 (141.1) & 29816 (149.1) \\
\textbf{MH}    & 28422 (142.1) & 27256 (136.3) & 33924 (169.6) & 26670 (133.4) & 27052 (135.3) & 28702 (143.5) \\
\textbf{Soc}   & 27116 (135.6) & 23353 (116.8) & 33834 (169.2) & 27850 (139.3) & 26997 (135.0) & 27074 (135.4) \\
\textbf{Tech}  & 28044 (140.2) & 27071 (135.4) & 33388 (166.9) & 26759 (133.8) & 28394 (142.0) & 26920 (134.6) \\
\hline
\end{tabular*}}
\caption{Number of labeled query-passage pairs and pairs per query (in parentheses) of each search task in DisastIR.}
\label{tab:qrels_stats}
\vspace{-5pt}
\end{table*}

\subsection{Assessment Candidate Pool Development}
\label{sec: candidate_pool}

Given the large size of the corpus, annotating all possible (query, passage) pairs is impossible \citep{thakur2021beir}. Following prior work, we construct a candidate pool for each query using existing retrieval models. Inspired by TREC's standard practice, where top-ranked passages from multiple systems are aggregated to form the candidate set, we adopt a similar strategy in DisastIR. 

Specifically, for each query, we collect the top $10$ retrieved passages from 30  retrieval models under two retrieval settings: exact and ANN search settings (detailed in Section~\ref{sec:experiment_steup}). These models also serve as baselines for performance evaluation, following practices in recent work \citep{rahmani2024syndl, wang2024feb4rag}. The candidate pool for each query is formed by taking the union of passages retrieved under both settings. 

\subsection{Relevance Labeling}
\label{sec: reve_label}

Once query-passage pairs are prepared, we annotate them using an LLM. Recent studies have shown that LLMs can reliably produce relevance judgments that align closely with human annotations \citep{rahmani2024synthetic, rahmani2024syndl, rahmani2025judging, wang2024feb4rag}. Furthermore, \citet{wang2024feb4rag, rahmani2024judgeblender} demonstrate that ensembling relevance scores from multiple prompts or LLMs yields more robust and calibrated annotations.

To this end, we design three diverse prompts for each search intent and use a single LLM to generate relevance scores. The prompts, inspired by \citet{thomas2024large, farzi2024best, rahmani2025judging}, are: (1) zero-shot direct scoring—a single-pass judgment; (2) chain-of-thought reasoning—a multi-step prompt mimicking human-style reasoning; and (3) multi-dimensional attribute scoring—relevance decomposed into interpretable sub-criteria. For each search intent, relevance is defined to align with its specific objectives, reflecting the varying interpretations of “relevance” across different task types \citep{dai2022promptagator}. Full prompt templates are provided in Appendix~\ref{appendix:prompts_qrels}.

Relevance scores are assigned on a 4-point scale (0 to 3) for all intents, except STS, which follows a 6-level scale as in \citet{agirre2013sem, cer2017semeval}. The final score for each pair is computed by averaging scores from three prompts.

\begin{table}[ht]
\centering
\small
\renewcommand{\arraystretch}{1.1}
\setlength{\tabcolsep}{5pt}
\begin{tabular}{lccccc}
\hline
 & \textbf{Count} & \textbf{Avg} & \textbf{Median} & \textbf{Min} & \textbf{Max} \\
\hline
Query   & 9,600     & 33.75     & 19  & 2   & 281 \\
Passage & 239,704   & 197.17    & 224 & 6   & 2{,}536 \\
\hline
\end{tabular}
\caption{Statistics of number of query and passage and their token lengths. Tokenization is based on the \texttt{cl100k\_base} tokenizer (used in GPT-4 / GPT-3.5).}
\label{tab:query_passage_length}
\vspace{-5pt}
\end{table}

\begin{table*}[ht]
\centering
\small
\renewcommand{\arraystretch}{1.15}
\setlength{\tabcolsep}{4pt}
\begin{tabular*}{\textwidth}{@{\extracolsep{\fill}}@{\hspace{5pt}}l|c|c|cccccc|ccc@{\hspace{5pt}}}
\hline
\textbf{Model} & \textbf{Param} & \textbf{Size} & \multicolumn{6}{c|}{\textbf{Exact} $\uparrow$} & \textbf{Ex.} & \textbf{ANN} & \textbf{Drop} \\
\textbf{Name}  & \textbf{Size}  & \textbf{Bin}  & \textbf{QA}    & \textbf{QAdoc} & \textbf{TW} & \textbf{FC} & \textbf{NLI}   & \textbf{STS}   & \textbf{Avg}   & \textbf{Avg}   & \textbf{(\%)}   \\
\hline

Linq-Embed-Mistral                 & 7B    & XL     & 74.40 & \textbf{70.50} & 64.22 & \textbf{70.77} & 52.56 & 71.35 & \textbf{67.30} & \textbf{66.98} & 0.48 \\

SFR-Embedding-Mistral              & 7B    & XL     & 71.50 & 67.34 & \textbf{69.62} & \underline{70.39} & 51.08 & 72.71 & \underline{66.71} & \underline{66.39} & 0.48 \\

inf-retriever-v1                   & 7B    & XL     & \underline{72.84} & 66.92 & \underline{66.37} & 65.76 & 52.02 & \underline{76.00} & 66.65 & 65.98 & 1.01 \\

inf-retriever-v1-1.5b              & 1.5B  & XL     & 69.47 & 64.40 & 63.08 & 65.49 & 54.14 & 73.96 & 65.09 & 64.85 & 0.37 \\
NV-Embed-v2                        & 7B    & XL     & \textbf{74.55} & \underline{69.51} & 42.55 & 68.39 & \textbf{58.39} & \textbf{76.13} & 64.92 & 64.57 & 0.54 \\

gte-Qwen2-1.5B-instruct            & 1.5B  & XL     & 69.96 & 59.21 & 65.21 & 62.84 & \underline{55.73} & 73.61 & 64.43 & 64.24 & 0.29 \\

multilingual-e5-large              & 560M  & Large  & 67.08 & 64.08 & 62.99 & 60.06 & 51.20 & 74.14 & 63.26 & 62.79 & 0.74 \\

e5-mistral-7b-instruct             & 7B    & XL     & 65.65 & 65.16 & 63.42 & 67.94 & 47.68 & 66.39 & 62.71 & 61.99 & 1.15 \\

multilingual-e5-large-instruct     & 560M  & Large  & 68.14 & 64.72 & 62.46 & 66.96 & 48.75 & 63.53 & 62.43 & 62.01 & 0.67 \\
e5-small-v2                        & 33M   & Small  & 65.66 & 62.84 & 60.10 & 61.78 & 47.12 & 73.93 & 61.90 & 61.48 & 0.68 \\
e5-base-v2                         & 109M  & Medium & 65.54 & 62.91 & 57.76 & 62.11 & 45.52 & 73.73 & 61.26 & 60.72 & 0.88 \\
e5-large-v2                        & 335M  & Large  & 60.03 & 63.24 & 55.48 & 62.03 & 50.96 & 74.09 & 60.97 & 60.45 & 0.85 \\
NV-Embed-v1                        & 7B    & XL     & 68.14 & 62.87 & 56.13 & 59.85 & 48.25 & 67.11 & 60.39 & 59.60 & 1.31 \\
granite-embedding-125m             & 125M  & Medium & 64.63 & 60.85 & 46.55 & 62.56 & 48.11 & 71.06 & 58.96 & 58.60 & 0.61 \\
gte-Qwen2-7B-instruct              & 7B    & XL     & 70.30 & 47.65 & 63.24 & 31.87 & 53.88 & 74.86 & 56.97 & 55.99 & 1.72 \\
snowflake-arctic-embed-m-v2.0      & 305M  & Medium & 61.28 & 62.31 & 47.20 & 57.84 & 42.43 & 64.56 & 55.94 & 55.15 & 1.41 \\

mxbai-embed-large-v1               & 335M  & Large  & 64.37 & 62.79 & 40.07 & 58.30 & 40.26 & 67.96 & 55.62 & 55.25 & 0.67 \\

gte-base-en-v1.5                   & 137M  & Medium & 60.46 & 55.85 & 46.44 & 52.34 & 39.85 & 70.41 & 54.22 & 53.93 & 0.53 \\

bge-base-en-v1.5                   & 109M  & Medium & 51.65 & 52.89 & 46.78 & 60.13 & 41.41 & 68.56 & 53.57 & 53.13 & 0.82 \\

gte-large-en-v1.5                  & 434M  & Large  & 67.46 & 58.37 & 39.71 & 52.90 & 34.79 & 66.51 & 53.29 & 53.21 & 0.15 \\

snowflake-arctic-embed-l-v2.0      & 568M  & Large  & 55.20 & 59.29 & 38.26 & 60.23 & 41.23 & 62.64 & 52.81 & 52.32 & 0.93 \\

bge-large-en-v1.5                  & 335M  & Large  & 56.88 & 54.56 & 32.32 & 55.03 & 35.25 & 64.43 & 49.74 & 49.04 & 1.41 \\

bge-small-en-v1.5                  & 33M   & Small  & 56.87 & 51.24 & 25.19 & 55.30 & 32.95 & 64.46 & 47.67 & 47.00 & 1.41 \\

snowflake-arctic-embed-s           & 33M   & Small  & 38.69 & 28.82 & 21.43 & 47.30 & 40.02 & 66.95 & 40.54 & 38.15 & 5.90 \\

snowflake-arctic-embed-m-v1.5      & 109M  & Medium & 25.66 & 30.43 & 18.09 & 48.10 & 42.98 & 64.20 & 38.24 & 36.85 & 3.63 \\

snowflake-arctic-embed-l           & 335M  & Large  & 40.73 & 30.33 & 15.11 & 32.60 & 34.44 & 56.11 & 34.89 & 32.17 & 7.80 \\


thenlper-gte-base                  & 109M  & Medium &  9.16 &  5.34 & 38.06 & 60.58 & 42.80 & 45.99 & 33.66 & 32.22 & 4.28 \\

snowflake-arctic-embed-m           & 109M  & Medium & 33.26 & 14.22 &  8.62 & 35.16 & 38.75 & 56.21 & 31.02 & 29.42 & 5.16 \\

snowflake-arctic-embed-m-long      & 137M  & Medium & 21.43 & 10.84 &  19.49 & 36.20 & 41.90 & 55.00 & 30.81 & 29.30 & 4.90 \\

thenlper-gte-small                 & 33M   & Small  &  18.20 &  9.08 & 11.04 & 49.81 & 37.71 & 55.47 & 30.22 & 29.43 & 2.61 \\

\hline
\end{tabular*}
\caption{Performances of 30 evaluated IR models in DisastIR. Models are ranked by their overall performance under exact search (highest to lowest) in DisastIR. ``Size Bin'' indicates its model parameter size bin category (small, medium, large, and extra large as defined in Appendix \ref{appendix:eva_models}). ``TW'' represents Twitter. Overall performance across all queries under exact and ANN search are in ``Ex. Avg'' and ``ANN Avg'' columns. ``Drop'' shows the percentage decrease from exact to ANN average scores. \textbf{Bold} indicates the highest value, and \underline{underline} indicates the second-highest. E5-small, base, large-v2, and granite-embedding use knowledge distillation during fine-tuning, which involves additional training signals. Performances across different event types are shown in Table~\ref{tab:model_eventtype_ranking_exact}.
}
\label{tab:model_searchtask_ranking_with_bin}
\vspace{-5pt}
\end{table*}

\section{DisastIR Benchmark Analysis}

\subsection{Query and Passage Characteristics}

\paragraph{Query and Passage Lengths.} As shown in Table~\ref{tab:query_passage_length}, the average query length is 33.75 tokens, with a median of 19, and a long tail extending to 281 tokens. This variation reflects the diversity of search intents, from short entity-style queries to detailed information needs typical in real-world disaster management scenarios. Passages are much longer on average (197.17 tokens), with a median of 224, and some exceeding 2,500 tokens. This wide distribution captures the diversity of disaster management-related texts, including both brief updates and detailed descriptions like event summaries or emergency protocols.

The corpus comprises both original and synthetic passages, with synthetic passages making up only 6.8 \% of the total. They come from two sources: 8,000 passages generated from page content (~3.3 \%) and 8,464 document-level summaries serving as proxies for full documents (~3.5 \%). Synthetic passages are introduced to enhance diversity and support varied search intents. Original passages, extracted from PDFs, are formal in style and, when chunked, often miss the broader document context. Some search scenarios, such as Twitter retrieval, require informal text featuring emojis, hashtags, or colloquial expressions, which are largely absent in PDF text. Other scenarios, such as QAdoc, demand whole-document understanding, where LLM-generated summaries provide an effective substitute since full texts typically exceed the input limits of IR models. Evaluation of LLM-generated passages is validated in Appendix \ref{appendix:LLM_passage_page}.

\paragraph{Labeled Query-Passage Pairs.} Table~\ref{tab:qrels_stats} summarizes the distribution of labeled query-passage pairs. In total, we obtained 1,341,986 labeled pairs, with each query linked to an average of 140 passages. 

As shown in Table~\ref{tab:qrels_stats}, Twitter-related search tasks tend to have a higher average number of query-passage pairs per query. The candidate pool for each query is built by merging the top 10 passages retrieved by 30 different models. This larger pool in Twitter tasks suggests greater divergence in model outputs, indicating lower agreement among retrieval models when ranking passages in social media contexts within disaster management scenarios. Additional analyses of labeled query-passage pairs are provided in Appendix~\ref{appendix:additional_labeled_qa_pairs}.

\subsection{LLM-based vs. Human Labeling}
\label{sec: llm_human_label}

Since relevance scores in DisastIR are judged by LLM, it is vital to evaluate their consistency with human annotations. Thus, we construct the LVHL dataset (\textbf{L}LM-based \textbf{V}s. \textbf{H}uman \textbf{L}abeling) by sampling disaster management-related query-passage pairs with human-labeled relevance scores from several open-source datasets. MS MARCO \citep{bajaj2016ms} and TriviaQA \citep{joshi2017triviaqa} are for QA, ALLNLI \citep{allnli_huggingface} and XNLI \citep{conneau2018xnli} for NLI, Climate-Fever \citep{diggelmann2021climate} for FC, and STSB \citep{cer2017semeval} for STS. Appendix~\ref{appendix:LVHL} provides details on the construction of LVHL.

The LLM-based relevance scores for each query-passage pair in LVHL are computed as described in Section~\ref{sec: reve_label}. Since most human-annotated relevance scores in LVHL are binary, we follow \citet{wang2024feb4rag} and binarize the LLM scores into two levels: relevant (score > 0) and not relevant (score = 0), to enable meaningful comparison. 

To assess agreement between LLM-based and human relevance labeling, we compute Cohen’s kappa for each search intent. All datasets yield kappa scores above 0.6 (Figure~\ref{fig:cohen_label}), with an average of 0.77, indicating substantial agreement. These suggest that LLM-generated relevance scores align well with human judgments and can reliably substitute for manual annotation in DisastIR.

We further conduct a controlled in-domain annotation experiment. Specifically, we sample 96 query–passage pairs from the DisastIR labeling pool, covering all 48 intent–event type combinations, and ask three PhD students specializing in disaster management to independently annotate them using the same multi-level relevance scale as in our LLM pipeline. Annotation guidelines are adapted from the chain-of-thought reasoning procedure employed for LLM labeling.

Inter-annotator agreement was substantial, with Fleiss’ kappa reaching 0.777. Comparing LLM-generated scores with human annotations yielded Cohen’s kappa values of 0.681–0.734 across individual annotators, 0.690 when averaged, and 0.803/0.715 under majority vote (minimum/maximum tie-breaking), again demonstrating substantial agreement (Table~\ref{tab:domain_kappa}). These findings confirm that LLM-based labeling is not only consistent with external human judgments (as shown in LVHL) but also robustly aligned with expert assessments within the disaster management domain.

\subsection{LLM vs. Human-generated User Query}
\label{sec:llm_human_query}

To evaluate whether LLM-generated queries can serve as a reliable alternative to human-authored ones for retrieval benchmarking, we construct LVHQ (\textbf{L}LM \textbf{V}s. \textbf{H}uman-generated \textbf{Q}uery), a comparison set spanning all 48 retrieval tasks. For each task, both an LLM-generated and a human-written query are created based on the same domain passage. All query-passage pairs are annotated using the same method as in DisastIR. Appendix~\ref{appendix:LVHQ} provides full details on the construction of LVHQ.

We evaluate all selected baseline models using LVHQ under exact search for both human- and LLM-generated queries (see Section~\ref{sec:experiment_steup} for evaluation setup up). Model performance, measured by NDCG@10, shows highly consistent results across the two query types, with a Kendall’s $\tau$ of 0.9264, indicating strong agreement in model evaluations.

\begin{figure*}[ht]
    \centering
    \includegraphics[width=\textwidth]{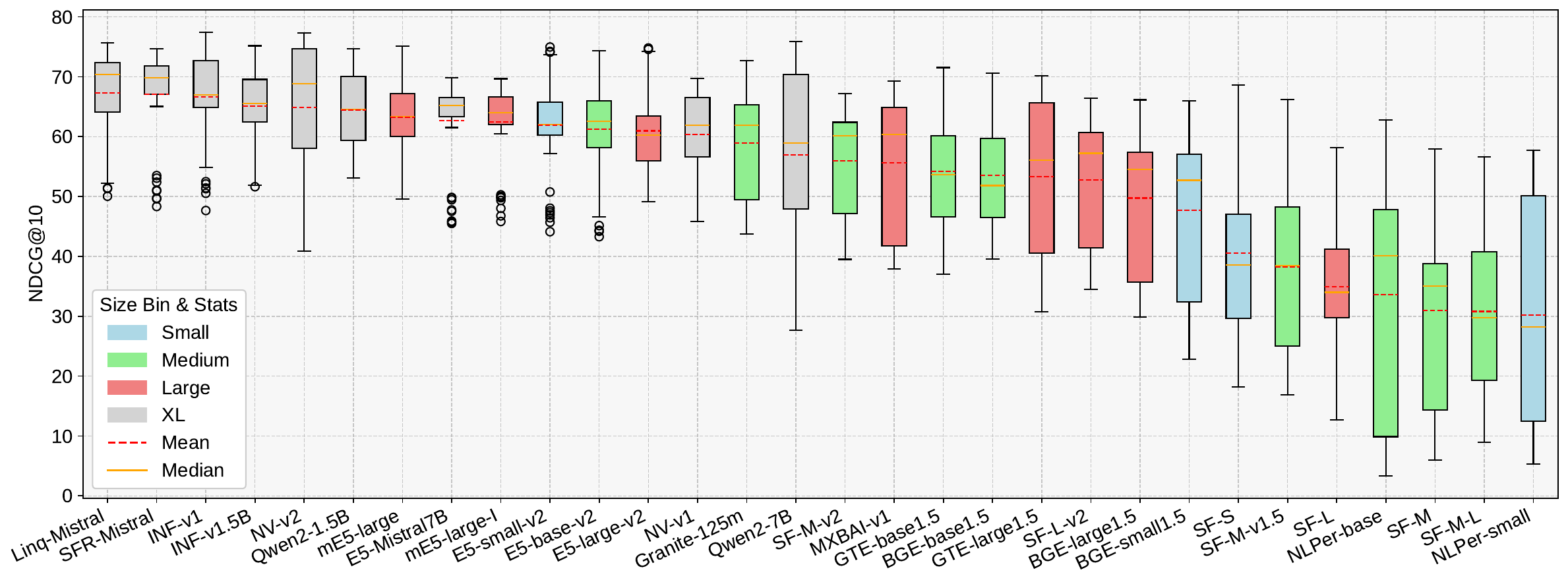}
    \caption{Distribution of evaluated models' performances across all 48 tasks. The full name of each model in the X axis is listed in Model Name column in Table \ref{tab:model_searchtask_ranking_with_bin}.}
    \label{fig:box_plot}
\vspace{-10pt}
\end{figure*}

\section{Experimental Setup}
\label{sec:experiment_steup}

\subsection{Models}
\label{Models:model_selection}

DisastIR is adopted to comprehensively evaluate open-source IR models and support the selection of suitable IR models for real-world disaster management applications. Models are chosen based on two criteria: (1) strong performance on the MTEB retrieval benchmark; and (2) inclusion in widely adopted embedding model families such as \texttt{BGE} \citep{chen2024m3, xiao2024c}, \texttt{E5} \citep{wang2022text, wang2023improving}, \texttt{Snowflake Arctic} \citep{merrick2024embedding}, and \texttt{GTE} \citep{li2023towards, zhang2024mgte}, which are commonly used as baselines and in downstream IR tasks \citep{sun2024mair, xu2024bmretriever, lee2024gecko, lee2024nv, cao2025enhancing, park2025mirage}.

We select 30 models with parameter sizes ranging from 33 million to 7 billion.
Detailed descriptions of these models and their implementations are provided in Appendix~\ref{appendix:eva_models}.

\subsection{Evaluation}
\label{Models:evaluation}

We evaluate model performance under two retrieval settings, exact and ANN, using Normalized Discounted Cumulative Gain at rank 10 (NDCG@10) as the primary metric, consistent with prior works.


\paragraph{(1) Exact Brute-force Retrieval.} Following prior work such as BEIR \citep{thakur2021beir}, InstructIR \citep{instructir2023}, FollowIR \citep{weller2024followir}, and MAIR \citep{zhang2024mair}, we compute similarity scores between each user query and all passages in the corpus, retrieving the top $k$ most similar ones.  This setting reflects model performance under ideal retrieval conditions.


\paragraph{(2) Approximate Nearest Neighbor (ANN) Retrieval.}
For large-scale corpora, brute-force retrieval is computationally infeasible. A common solution is a multi-stage architecture, where an ANN search retrieves a candidate set of passages, which are then re-ranked for final output \citep{tu2020approximate, macdonald2021approximate}. To reflect real-world large-scale disaster information retrieval scenarios, we also evaluate model performance during the candidate generation stage using ANN search. We adopt 
the HNSW (hierarchical navigable small world) algorithm \citep{malkov2018efficient}, to retrieve top $k$ passages per query using precomputed embeddings. For fair comparison, $k$ is set to match the value used in exact search.

\begin{figure}[!htbp]
    \centering
    \includegraphics[width=\linewidth]{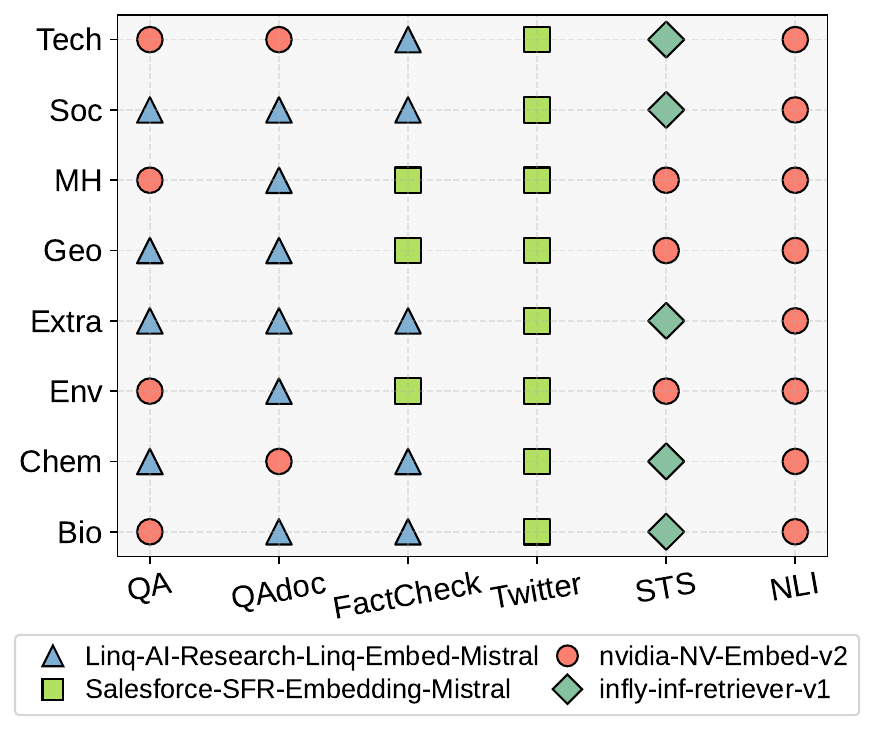}
    \caption{Best-performing models in each search task.}
    \label{three_best_dis}
\end{figure}

\begin{figure*}[!ht]
    \centering
    \includegraphics[width=\textwidth]{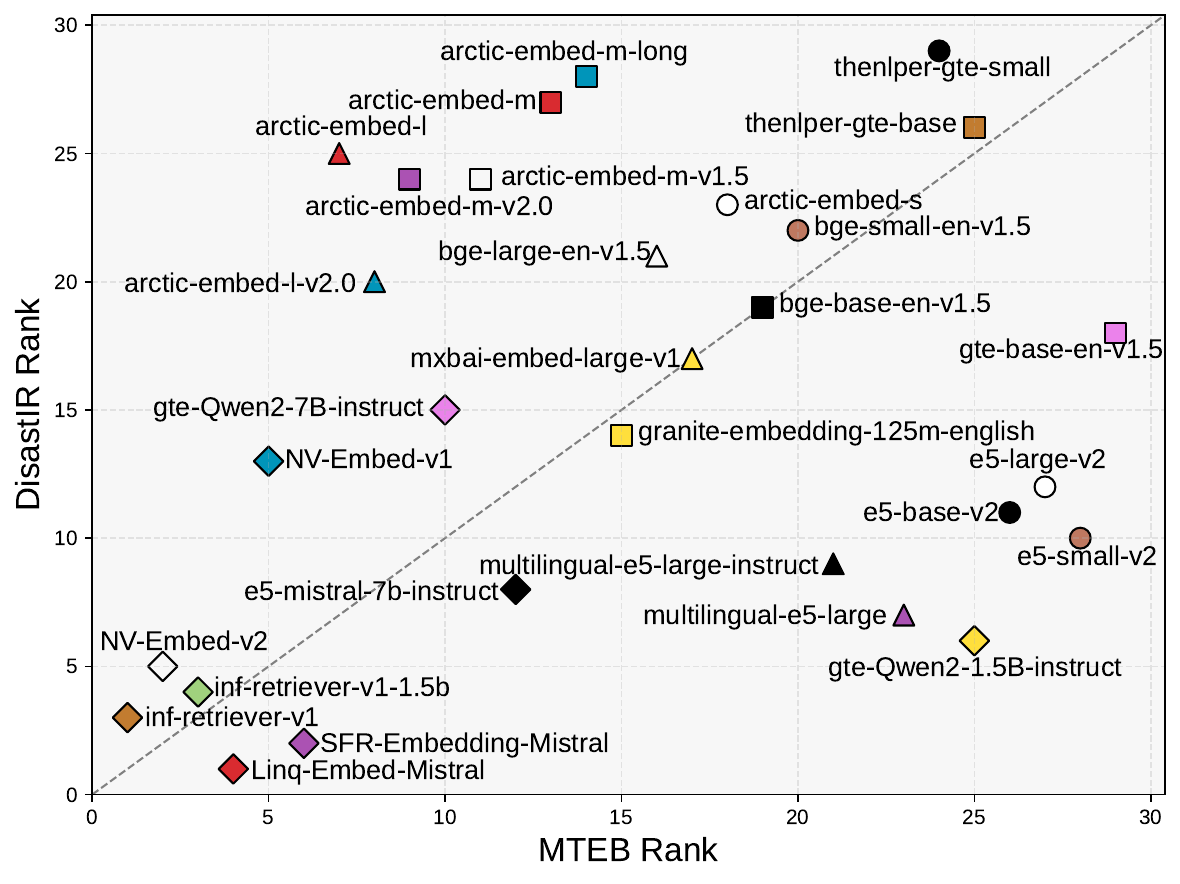}
    \caption{Comparison between DisastIR and MTEB model rankings. Legend shapes indicate the model size bin:
  $\blacklozenge$ XL, 
  $\blacktriangle$ Large, 
  $\blacksquare$ Medium, 
  $\bullet$ Small.}
    \label{fig:disas-mteb}
\vspace{-10pt}
\end{figure*}

\section{Evaluation Results}

\subsection{Overall Performance}
\label{sec: overall_performance}

Table~\ref{tab:model_searchtask_ranking_with_bin} summarizes the overall performance of all 30 evaluated models across all queries in DisastIR, with detailed results for each search task provided in Appendix~\ref{appendix:model_performance}. The  \texttt{Linq-Embed-Mistral} model achieves the best performance under exact and ANN search settings, followed closely by \texttt{SFR-Embedding-Mistral} (0.877\% and 0.881\% lower, respectively).

Among all non-XL models, multilingual-e5-large performs best, reaching 94.0\% and 93.00\% of the top model's performance. Notably, the lightweight \texttt{e5-small-v2} model (33M parameters) achieves 91.98\% and 91.17\% of the top model’s performance, despite being 212 times smaller in size. The E5-V2 series \citep{wang2022text} and Granite-Embedding \citep{awasthy2025granite} leverage knowledge distillation during fine-tuning, achieving performance that surpasses all models of the same scale and even outperforms many substantially larger models. This highlights the effectiveness of knowledge distillation in enhancing the performance of smaller models.

The \texttt{Snowflake-arctic-embed-l} model shows the largest performance drop (7.80\%) under ANN search compared to exact search (Table \ref{tab:model_searchtask_ranking_with_bin}). Most models exhibit drops within 2\%; only five exceeded this margin, four of which belong to snowflake family, indicating strong robustness when switching from exact to ANN search in DisastIR. 
All subsequent analyses are based on exact search; analyses under ANN search can be conducted similarly. 

\subsection{Performance across all 48 Tasks}
\label{section: perform_48_tasks}

Figure~\ref{fig:box_plot} presents the performance distribution of all evaluated models across all 48 search tasks. All top-5 performance models show great variability across tasks, as reflected by the large interquartile range (IQR). This highlights the limited cross-task robustness of current general domain retrieval models and underscores the need to design methods that enhance cross-task consistency, rather than optimizing solely for higher average performance.

As shown in Figure~\ref{three_best_dis}, \textbf{no single model consistently outperforms others across all 48 tasks}. Instead, top performance is distributed among four models: Linq-Embed-Mistral, inf-retriever-v1, SFR-Embedding-Mistral, and NV-Embed-v2.  \textbf{This highlights the complexity and diversity of disaster management-related retrieval tasks and reinforces the need for domain-specific IR models in real-world disaster management scenarios.} Appendix \ref{appendix:additional_48_tasks_analyses} provides additional analyses of model performance across 48 tasks.

Among these top-performing models, only NV-Embed-v2 is accompanied by a public technical report~\citep{lee2024nv}, enabling a closer look at its varying performance across intents. Its training data emphasizes fact-checking, NLI, and QA but excludes Twitter, explaining strong results on QA and NLI and weaker performance on Twitter. Template usage further matters: NV-Embed-v2 benefits from alignment between training and inference for QA, QAdoc, FC, and NLI, but lacks such alignment for Twitter. Architecturally, unlike models relying on the final <EOS> token, NV-Embed-v2 employs a latent attention layer, which helps it achieve the highest average NDCG@10 (69.39) among top models when Twitter is excluded.


\subsection{Comparison with General Domain}
\label{sec:compare_with_mteb}

Figure~\ref{fig:disas-mteb} compares model rankings in DisastIR and MTEB. Ranking value of each model is based on overall performance in DisastIR and official retrieval scores from the MTEB English leaderboard. The Spearman correlation between the two rankings is 0.252 ($p = 0.188$), indicating no significant correlation. This suggests that \textbf{strong performance on general-domain benchmarks does not guarantee effectiveness in disaster management-related retrieval}. For example, models in \texttt{snowflake} family perform well in MTEB but poorly in DisastIR, while models from the \texttt{E5} family show the opposite trend.

Furthermore, when computational resources are limited and large models are impractical to serve, relying solely on MTEB rankings for model selection, such as choosing \texttt{snowflake-arctic-embed-l}, may fail to retrieve critical or relevant content. \textbf{These discrepancies underscore the necessity of a domain-specific benchmark like DisastIR to guide retrieval model selection across different disaster management-related search tasks.}

\section{Conclusion}

In this work, we introduce and publicly release DisastIR, the first comprehensive retrieval benchmark for evaluating model performance in disaster management contexts. DisastIR consists of 9,600 user queries and more than 1.3 million labeled query-passage pairs, spanning 48 retrieval tasks defined by six search intents and eight general disaster event types, covering 301 specific event types.

Using DisastIR, we evaluate 30 SOTA open-source retrieval models under both exact and ANN search settings. Our findings provide practical guidance for selecting appropriate IR models based on task type and computational constraints, supporting timely and effective access to critical information in disaster management scenarios.

\section*{Limitations}
While DisastIR represents a significant step toward domain-specific evaluation in disaster information retrieval, several aspects merit further enhancement. DisastIR currently focuses on English-language resources. Expanding DisastIR to multilingual settings would enable broader applicability. Furthermore, tables and figures in domain-specific PDF files may contain useful domain knowledge. Further study could consider extracting this critical information for evaluation set development.

\section*{Ethics Statement}
DisastIR is designed to support disaster management by improving the evaluation and selection of retrieval models. All data used in the benchmark are sourced from publicly available materials, and no personally identifiable information is included. All contents generated by LLMs are evaluated by a human expert to ensure no offensive content is included in the DisastIR. We recognize potential risks associated with the misuse of retrieval models in disaster contexts, such as the spread of disinformation during crises. To mitigate these risks, DisastIR is intended solely for evaluation purposes and is released for research use only.

\section*{Acknowledgments}
This work used ACES at TAMU, DeltaAI and Delta GPU at the National Center for Supercomputing Applications through allocation CIV250019 and CIV250021 from the Advanced Cyberinfrastructure Coordination Ecosystem: Services \& Support (ACCESS) program, which is supported by U.S. National Science Foundation grants \#2138259, \#2138286, \#2138307, \#2137603, and \#2138296.

\bibliography{custom}
\bibliographystyle{acl_natbib}

\newpage

\appendix

\section{Structural PDF File Processing Pipeline}
\label{appendix:pdf_pipeline}

\begin{table*}[!h]
\centering
\small
\renewcommand{\arraystretch}{1.15}
\setlength{\tabcolsep}{3pt}
\begin{tabular*}{\textwidth}{@{\extracolsep{\fill}}@{\hspace{5pt}}l|cccccccc@{\hspace{5pt}}}
\hline
\textbf{Model}
  & \textbf{Bio}
  & \textbf{Chem}
  & \textbf{Env}
  & \textbf{Extra}
  & \textbf{Geo}
  & \textbf{MH}
  & \textbf{Soc}
  & \textbf{Tech} \\
\hline
Linq-Embed-Mistral & \textbf{68.07} & \textbf{66.75} & \underline{67.82} & \textbf{67.30} & \underline{65.40} & \textbf{67.13} & \textbf{68.56} & \underline{67.36} \\

SFR-Embedding-Mistral & 67.74 & \underline{66.70} & \textbf{68.03} & \underline{66.43} & \underline{65.40} & \underline{67.03} & \underline{68.15} & \textbf{67.37} \\

inf-retriever-v1 & \underline{67.75} & 66.40 & 66.92 & 65.04 & \textbf{65.52} & 66.66 & \underline{68.15} & 66.75 \\

inf-retriever-v1-1.5b & 65.71 & 64.45 & 65.36 & 63.33 & 64.31 & 65.45 & 66.59 & 65.54 \\

NV-Embed-v2 & 65.23 & 64.51 & 65.55 & 64.79 & 63.51 & 64.94 & 65.95 & 64.89 \\

gte-Qwen2-1.5B-instruct & 65.15 & 64.35 & 64.94 & 62.56 & 63.16 & 63.97 & 66.12 & 65.16 \\

multilingual-e5-large & 63.00 & 62.91 & 64.01 & 62.11 & 62.59 & 63.10 & 64.26 & 64.10 \\

e5-mistral-7b-instruct & 63.33 & 62.13 & 63.49 & 62.25 & 60.88 & 62.70 & 63.85 & 63.04 \\

multilingual-e5-large-instruct & 62.64 & 61.79 & 62.76 & 62.48 & 60.49 & 62.70 & 63.60 & 62.95 \\

e5-small-v2 & 62.73 & 62.56 & 64.17 & 60.59 & 60.71 & 62.68 & 62.29 & 62.29 \\
e5-base-v2 & 61.86 & 61.34 & 63.05 & 60.97 & 61.08 & 61.96 & 62.98 & 61.75 \\
e5-large-v2 & 60.09 & 59.49 & 60.84 & 59.77 & 59.18 & 60.48 & 61.15 & 60.74 \\
NV-Embed-v1 & 61.37 & 58.86 & 61.91 & 58.84 & 59.13 & 61.13 & 61.23 & 60.66 \\
granite-embedding-125m & 58.00 & 57.52 & 58.09 & 57.21 & 56.80 & 57.89 & 58.32 & 57.61 \\

arctic-embed-m-v2.0 & 56.47 & 55.78 & 56.14 & 55.05 & 54.88 & 55.66 & 55.98 & 55.29 \\

mxbai-embed-large-v1 & 56.18 & 55.38 & 55.96 & 54.77 & 54.56 & 55.47 & 55.88 & 55.07 \\
gte-base-en-v1.5 & 54.78 & 54.29 & 55.49 & 53.19 & 52.99 & 54.19 & 54.60 & 53.98 \\
bge-base-en-v1.5 & 53.57 & 52.78 & 54.18 & 52.38 & 51.98 & 53.18 & 53.77 & 52.88 \\
gte-large-en-v1.5 & 63.23 & 62.54 & 63.86 & 62.32 & 61.92 & 63.05 & 64.10 & 63.07 \\
snowflake-arctic-embed-l-v2.0 & 53.86 & 53.09 & 54.44 & 52.78 & 52.36 & 53.58 & 53.96 & 53.17 \\

bge-large-en-v1.5 & 53.18 & 52.49 & 53.88 & 52.08 & 51.78 & 52.79 & 53.49 & 52.58 \\
gte-Qwen2-7B-instruct & 52.90 & 52.20 & 53.70 & 51.90 & 51.60 & 52.60 & 53.20 & 52.30 \\
bge-small-en-v1.5 & 45.05 & 44.45 & 35.75 & 44.91 & 44.14 & 45.23 & 45.95 & 44.73 \\

snowflake-arctic-embed-s & 40.91 & 40.51 & 43.60 & 38.26 & 38.71 & 40.15 & 41.65 & 40.50 \\

snowflake-arctic-embed-m-v1.5 & 42.30 & 42.79 & 44.14 & 40.49 & 41.20 & 41.99 & 42.47 & 42.19 \\

snowflake-arctic-embed-l & 35.15 & 35.05 & 37.77 & 34.28 & 32.53 & 33.81 & 35.92 & 34.59 \\


thenlper-gte-base & 35.84 & 35.28 & 33.35 & 32.19 & 31.19 & 32.71 & 34.83 & 33.84 \\
snowflake-arctic-embed-m & 30.72 & 31.00 & 32.51 & 31.40 & 29.45 & 30.71 & 31.62 & 30.74 \\
snowflake-arctic-embed-m-long & 30.82 & 30.67 & 32.16 & 30.77 & 29.44 & 30.00 & 31.87 & 30.75 \\
thenlper-gte-small & 31.92 & 31.53 & 28.83 & 30.25 & 27.72 & 30.26 & 29.96 & 31.27 \\

\hline
\end{tabular*}
\caption{Performance of 30 open-source retrieval models under the exact-search setting across different disaster event types. Each cell shows the \textbf{mean} NDCG@10 over the 6 search tasks for that event type. 
}
\label{tab:model_eventtype_ranking_exact}
\end{table*}

\begin{figure}[ht]
  \centering
  \includegraphics[width=\linewidth]{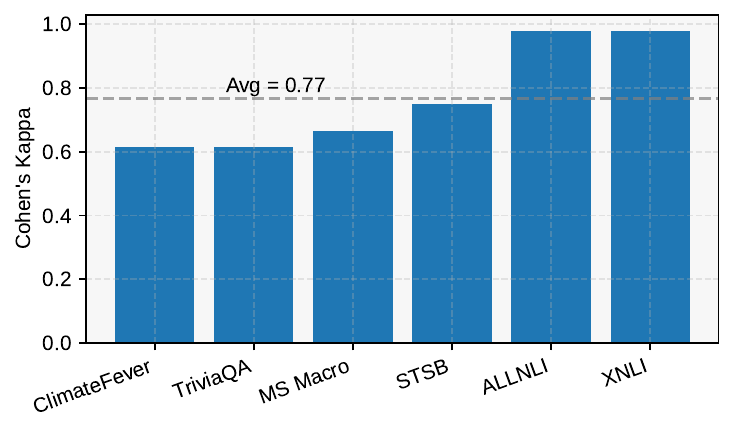}
  \caption{Cohen’s kappa scores between LLM-based and human-annotated relevance labels across all LVHL datasets, as described in Section~\ref{sec: llm_human_label}}
  \label{fig:cohen_label}
\end{figure}


All disaster management-related data (in PDF format) is obtained from publicly available sources with no personally identifiable information. Hence, explicit consent was not required. We chose PDFs because they typically contain more structured, information-rich, and credible content, often originating from peer-reviewed publications or official institutions. PDF files are collected using \texttt{googlesearch-python} (v1.3.0) and processed with \texttt{PyMuPDF} (v1.24.10) for content extraction. The extracted PDFs are then processed into text chunks through the following steps:

\paragraph{(1) Exact-URL Deduplication.} The URL of each downloaded PDF is recorded, and duplicate documents are removed by identifying identical download links. 
\paragraph{(2) Text Extraction and Preprocessing.} Each PDF file is converted into plain text, where tables and figures are removed following the work of \citep{wen2023chathome}.
\paragraph{(3) Locality-Sensitive Hashing (LSH) Deduplication.} After cleaning, we apply LSH-based near-duplicate detection to identify and remove documents with highly overlapping content.
\paragraph{(4) Semantic Chunking.}Cleaned documents are segmented into semantically coherent text chunks. Each chunk is constrained to fewer than 256 tokens to optimize retrievability while maintaining semantic integrity.
\paragraph{(5) Embedding-based Near Deduplication.}To further eliminate redundancy at the passage level, dense embeddings are computed for all chunks. An ANN index is built to retrieve the top-$k$ nearest chunks, and pairs with cosine similarity above 0.9 are removed.

Since our benchmark emphasizes retrieval performance across event types and search intents rather than source analysis, we summarize the corpus by the number of PDFs per general event type. Each general event type includes multiple specific event types, which were used as search queries. The number of collected PDFs is thus correlated with the number of associated specific types, as shown in Table \ref{tab:event-types-pdfs}.

\begin{table}[ht]
\centering
\small
\renewcommand{\arraystretch}{1.15}
\setlength{\tabcolsep}{4pt}
\begin{tabular*}{\columnwidth}{@{\extracolsep{\fill}}@{\hspace{5pt}}l|c@{\hspace{5pt}}}
\hline
\textbf{General event type} & \textbf{Number of PDFs (percentage)} \\
\hline
MH    & 1683 (19.89\%) \\
Bio   & 2343 (27.68\%) \\
Geo   & 868 (10.26\%) \\
Tech  & 1598 (18.88\%) \\
Extra & 254 (3.00\%) \\
Chem  & 767 (9.06\%) \\
Soc   & 234 (2.76\%) \\
Env   & 717 (8.47\%) \\
\hline
\end{tabular*}
\caption{Distribution of event types by number of PDFs}
\label{tab:event-types-pdfs}
\end{table}

\section{Evaluation of LLM-generated passages}
\label{appendix:LLM_gen_passage_eva}

\subsection{Evaluation of LLM document summaries}
\label{appendix:LLM_doc_summary}

\paragraph{Experiment design.}  
We evaluate whether each LLM-generated summary accurately captured the main content of the original material, which motivated their inclusion in the corpus. A total of 48 summaries (six per general event type) are randomly sampled. Three PhD students independently annotate each summary for fluency and content accuracy, following detailed guidelines adapted from \citet{song2024finesure}.

\paragraph{Results and analysis.}  
\begin{table}[h]
\centering
\end{table}

Fluency is defined as the percentage of summaries judged grammatically and semantically well-formed, while content accuracy measures whether summaries captured the main ideas of the source material. The average results show 100\% fluency and 96.88\% content accuracy. The Fleiss’ Kappa score of 0.793 indicated substantial inter-annotator agreement, supporting the reliability of the evaluation. Overall, LLM-generated summaries are fluent and highly faithful to the original documents.

\subsection{Evaluation of LLM-generated passages from page content}
\label{appendix:LLM_passage_page}

\paragraph{Experiment design.}  
We further assess LLM-generated passages created directly from page content, focusing on their fluency and adherence to style requirements. The QAdoc intent is excluded, as its format (document summaries) had already been validated. We randomly sampled 40 passages (eight per intent across five intents) and asked three PhD students to evaluate them. The evaluators followed the same style guidelines provided to the LLM during generation.

\paragraph{Results and analysis.}  
\begin{table}[h]
\centering
\end{table}

LLM-generated passages achieved an average fluency score of 98.75\% and a style compliance score of 92.5\%. The Fleiss’ Kappa of 0.760 again confirmed substantial agreement among annotators. These results indicate that the generated passages are both fluent and well aligned with intent-specific stylistic requirements.

\subsection{Role of noisy passages in realistic IR evaluation}
\label{appendix:LLM_passage_noise}

While our evaluations confirm the overall quality of LLM-generated content, we emphasize the value of including imperfect or noisy passages in the corpus. Real-world IR corpora naturally contain irrelevant or off-topic data, and such noise can enhance the realism of evaluations. Introducing noisy passages allows us to test the robustness of IR models by assessing whether they can correctly identify irrelevant content and assign low relevance scores. For example, a generated tweet unrelated to a query serves as a negative case; an effective IR model should detect this mismatch and rank it accordingly.

\section{Prompt Templates for Query Generation}
\label{appendix:prompts_query_gen}

Prompts for query generation based on disaster management-related passage under different search intents for QA, QAdoc, Twitter, FC, NLI, STS are in Tables \ref{tab:QA_query_prompt}, \ref{tab:QAdoc_query_prompt}, \ref{tab:Twitter_query_prompt}, \ref{tab:factcheck_prompt_query}, \ref{tab:nli_prompt_query}, and \ref{tab:STS_prompt_query}. 

\section{Prompt Templates for Relevance Labeling}
\label{appendix:prompts_qrels}

This section presents the prompt templates used for LLM-based relevance judgments across six search intents, employing three prompting strategies: Zero-shot Direct Scoring, Chain-of-Thought Decomposed Reasoning, and Multi-Dimensional Attribute Scoring. For QA, QAdoc, and Twitter tasks, we adapt templates from \citet{thomas2024large, farzi2024best}, as shown in Table~\ref{tab:QA_eva_prompt_template}. Based on these templates, we design relevance prompts for FC, NLI, and STS tasks, shown in Tables~\ref{tab:factcheck_prompt_templates}, \ref{tab:nli_prompt_templates}, and \ref{tab:sts_prompt_template}, respectively. For STS, we adopt only Zero-shot Direct Scoring, as our preliminary experiments show it yields higher agreement with human labels (Cohen’s kappa). The estimated cost of generating 9,600 user queries and labeling over 1.3 million query-passage using GPT-4o-mini API is about \$1,400.

\section{Additional Analyses of Labeled query-passage pairs}
\label{appendix:additional_labeled_qa_pairs}

As shown in Table~\ref{tab:score_distribution}, in certain retrieval tasks, such as NLI\_Bio, NLI\_Geo, the number of query-passage pairs assigned the highest relevance score is smaller than the number of user queries. This indicates that some queries do not have any passage in their candidate pool that is judged as fully relevant. For each query, we prompt an LLM to generate a directly relevant passage based on the associated domain passage and include it in the labeling pool (Section \ref{sec: user_query_gen}). However, the labeling results in Table~\ref{tab:positive_doc_score_dist} show that not all generated passages are considered fully relevant. This suggests that, even when guided by task-specific prompts, LLMs may produce passages that only partially address the query or fail to capture its key intent. 

Many recent works have tried to employ LLM to generate synthetic training data to improve the quality of retrievers \citep{wang2023improving, rajapakse2023improving, xu2024bmretriever, lee2024gecko}. This finding underscores the importance of consistency filtering \citep{alberti2019synthetic} to improve retrieval models' performance, as LLM will generate irrelevant pairs. This aligns with prior research highlighting the need for consistency filtering when leveraging LLM-generated data to train retrievers \cite{dai2022promptagator, xu2024bmretriever, lee2024gecko}.

\section{LVHL Dataset Construction}
\label{appendix:LVHL}

We use the names of 301 specific disaster event types as queries to search for disaster management-related user queries within each selected open-source dataset listed in Table~\ref{tab:dataset-overview}. For each dataset, we first filter queries by keyword matching and then prompt an LLM to further remove queries that are irrelevant to disaster management. From the remaining queries, we randomly select up to 400 queries per dataset. The corresponding passage and relevance score in each source dataset are also included. This process results in the final query-passage pairs along with the human-annotated relevance scores used in the LVHL dataset for evaluating the agreement of LLM-based and human-annotated relevance scores. 
\footnote{LVHL is used solely to evaluate agreement between LLM and human annotations. It is not suitable for benchmarking retrieval models in the disaster management area, as most queries are drawn from training sets of the source datasets.}

\section{LVHQ Dataset Construction}
\label{appendix:LVHQ}

We sample 48 domain passages developed in Section~\ref{sect: domain_knowledge_chunk}, ensuring one passage per retrieval task and keeping all sampled passages different from those used in developing DisastIR. For each passage, a domain expert in the disaster management field is asked to read the passage and write a realistic user query that reflects a practical information need based on the content, resulting in 48 human-authored queries. The Human expert is given the same instructions for the query written (shown in Tables \ref{tab:QA_query_prompt}, \ref{tab:QAdoc_query_prompt}, \ref{tab:Twitter_query_prompt}, \ref{tab:factcheck_prompt_query}, \ref{tab:nli_prompt_query}, \ref{tab:STS_prompt_query}) as those given to LLM to ensure fair comparison.  
In parallel, for the same set of passages, we also generate 48 queries using LLM in the same way as described in Section~\ref{sec: user_query_gen}. 

Each query, both human-authored and LLM-generated, is used to retrieve relevant passages from DisastIR corpus. As we have validated the agreement of LLM-based and Human-annotated relevance score in Section \ref{sec: llm_human_label}, all query-passage pairs are labeled in the same way as described in Section \ref{sec: candidate_pool} and Section \ref{sec: reve_label}. 

\begin{figure*}[ht]
    \centering
    \includegraphics[width=1.01\textwidth]{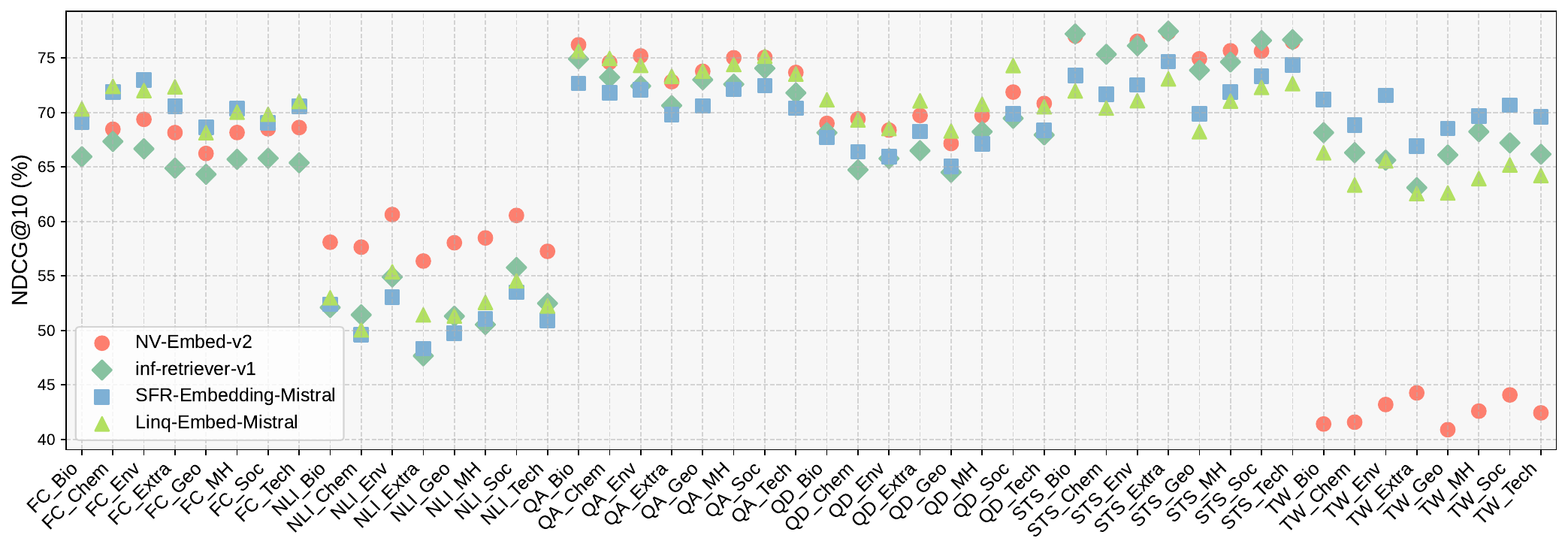}
    \caption{Performance of three top models across different tasks}
    \label{fig:three_best}
\end{figure*}

\begin{figure}[!htbp]
    \centering
    \includegraphics[width=1.01\linewidth]{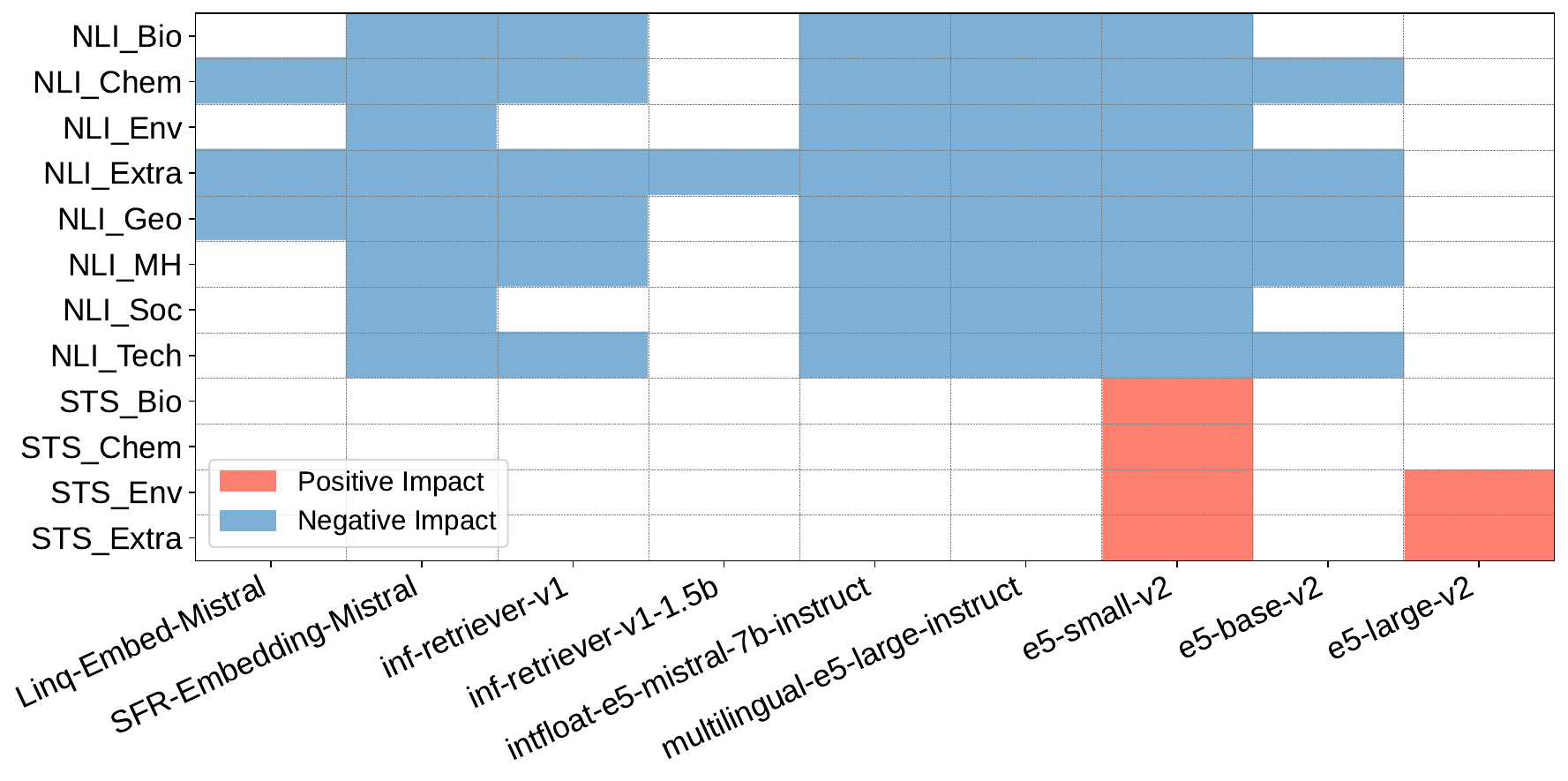}
    \caption{Distribution of outliers of evaluated models' performances}
    \label{Outliers}
\end{figure}

\section{Information of Evaluated Models and Model Implementation}
\label{appendix:eva_models}

Detailed information on all selected models is summarized in Table \ref{tab:model_specs}. The HuggingFace links and licenses of these models are in Table \ref{tab:model_links}. The model parameter size is categorized as four levels: small (<109M), medium (109M - 305M), large (305M- 1B), and extra large (XL) (> 1B). 

For each model, we follow official implementation guidelines to generate normalized query and passage embeddings. All evaluations are conducted in a zero-shot setting, with input sequences truncated to 512 tokens and a task-specific instruction prepended to each query. All models are run on a single NVIDIA A6000 GPU using HuggingFace Transformers, following the configurations specified in the official implementations.

\section{Performance of Evaluated Models}
\label{appendix:model_performance}

Performance of all evaluated models in all 48 search tasks in DisastIR is shown in Tables \ref{tab:baseline_models_group1}, \ref{tab:baseline_models_group2}, \ref{tab:baseline_models_group3}, \ref{tab:baseline_models_group4}, and \ref{tab:baseline_models_group5}. 

\section{Additional Analyses of Model Performance across 48 Tasks}
\label{appendix:additional_48_tasks_analyses}

\texttt{NV-Embed-v2} achieves the best performance on all NLI-related tasks (See Table \ref{tab:model_searchtask_ranking_with_bin} and Figure \ref{three_best_dis} in the main content). However, as shown in Figure~\ref{fig:three_best}, its poor results on Twitter-related tasks significantly lower its overall performance in DisastIR. This reflects its limitation in handling informal, noisy, and contextually ambiguous nature of social media content. Given the importance of Twitter as a real-time, crowd-sourced information source during disasters \citep{alam2021crisisbench, yin2024crisissense, lei2025harnessing}, this weakness raises concerns about its reliability in real-world disaster response scenarios.

All four models perform poorly on NLI-related tasks, with the best achieving only an average score of 58.39 (Figure~\ref{fig:three_best}). Further analysis of outliers in the box plot (See Figure~\ref{fig:box_plot} in the main content) reveals that tasks causing significant performance drops consistently involve NLI search intents (Figure~\ref{Outliers}). This reveals a key limitation of current open-source SOTA retrievers, that they struggle with the complex reasoning required for NLI tasks in disaster contexts. Such limitations may lead to incorrect results or failure to retrieve critical information, which can negatively impact decision-making in disaster situations.

\begin{table}[ht]
\centering
\small
\renewcommand{\arraystretch}{1.15}
\setlength{\tabcolsep}{4pt}
\begin{tabular*}{\columnwidth}{@{\extracolsep{\fill}}@{\hspace{5pt}}l|c@{\hspace{5pt}}}
\hline
\textbf{Reference Labeling} & \textbf{Cohen’s $\kappa$} \\
\hline
Annotator 1 & 0.695 \\
Annotator 2 & 0.681 \\
Annotator 3 & 0.734 \\
Rounded average & 0.690 \\
Majority vote (min in tie) & 0.803 \\
Majority vote (max in tie) & 0.715 \\
Fleiss’ $\kappa$ (inter-annotator) & 0.777 \\
\hline
\end{tabular*}
\caption{Agreement between LLM-based relevance labels and human annotations on the in-domain sample (96 query–passage pairs).}
\label{tab:domain_kappa}
\end{table}

\begin{table*}[ht]
\centering
\small
\renewcommand{\arraystretch}{1.2}
\setlength{\tabcolsep}{4pt}
\begin{tabular*}{\textwidth}{@{\extracolsep{\fill}}p{\textwidth}@{}}
\hline
\textbf{Information Need Generation Stage}\\
\hline
"Brainstorm a list of useful text retrieval tasks where the goal is: Given a user question, retrieve passages that directly answer the question. Here are a few examples: Given a question about evacuation procedures during a flood, retrieve a passage that explains the steps involved. Given a question about the cause of infrastructure failure in a disaster, retrieve a passage identifying the cause. Given a question about relief funding timelines, retrieve a passage providing the relevant information. Guidelines: Each task description should be one sentence that clearly describes the user question and the kind of answer passages to be retrieved. Focus on real-world domains like disaster planning, relief logistics, early warning systems, community impact, government response, etc. Your output should be a JSON list of 3 strings, each describing a distinct and useful text retrieval task. Only output the list. Be creative. No explanations or additional content."\\
\hline
\textbf{User Query Generation Stage}\\
\hline
"You have been assigned a retrieval task: \verb|{task}|. Your mission is to write one text retrieval example for this task in JSON format. The JSON object must contain the following keys: \\
\;\;- user\_query: a string, a random user search query specified by the retrieval task. \\
\;\;- positive\_document: a string, a relevant document for the user query. \\
Please adhere to the following guidelines: \\
\;\;- The user\_query should be \verb|{query_length}|, \verb|{clarity}|. \\
\;\;- All documents must be created independent of the query. Avoid copying the query verbatim. It is acceptable if some parts of the positive\_document are not topically related to the query. \\
\;\;- All documents should be \verb|{num_words}| long. \\
\;\;- Do not provide any explanation in any document on why it is relevant or not relevant to the query. The query and documents must be realistic and inspired by real-world content (e.g., disaster management). All generated content should be in English no matter the provided content language is. \\
\;\;- Both the query and documents require \verb|{difficulty}| level education to understand. \\
Your output must always be a JSON object only, do not explain yourself or output anything else. Be creative!"\\
\hline
\end{tabular*}
\caption{Prompt templates for user query generation in QA-related tasks. The \texttt{clarity} placeholder takes values: clear, understandable with some effort, and ambiguous. The \texttt{difficulty} placeholder includes: elementary school, high school, college, and PhD. For \texttt{query\_length}, possible values are: less than 10 words, 5 to 20 words, less than 20 words, at least 50 words, and at least 150 words. The \texttt{num\_words} placeholder takes values such as: at least 100 words, at least 200 words, at most 50 words, and 50 to 150 words.}
\label{tab:QA_query_prompt}
\end{table*}

\begin{table*}[ht]
\centering
\small
\renewcommand{\arraystretch}{1.2}
\setlength{\tabcolsep}{4pt}
\begin{tabular*}{\textwidth}{@{\extracolsep{\fill}}p{\textwidth}@{}}
\hline
\textbf{Information Need Generation Stage}\\
\hline
"Brainstorm a list of document retrieval tasks, where the goal is: Given a user query, retrieve documents that provide useful and relevant answers. Here are a few examples to get you started: Given a query about emergency evacuation procedures, retrieve a document that outlines the proper steps. Given a query asking how heatwaves affect public health, retrieve a document discussing the medical or environmental impacts. Given a query about funding for post-disaster recovery, retrieve a document describing the financial aid process. Given a query on how early warning systems reduce disaster risk, retrieve a document explaining their function and benefits. Guidelines: Each task should be a single sentence describing what the query is and what kind of document should be retrieved in response. Tasks should span a broad range of information needs, from facts to procedures to causal relationships. Focus on disaster management-related themes such as risk mitigation, emergency logistics, climate impact, institutional roles, and infrastructure damage. Your output should be a JSON list of 20 strings, each describing a distinct and useful text retrieval task. Only output the list. No explanations."\\
\hline
\textbf{User Query Generation Stage}\\
\hline
"You have been assigned a document retrieval task: \verb|{task}|. Your mission is to write one example for this task in JSON format. The JSON object must include: \\
\;\;- \texttt{user\_query}: a single, well-formed, natural language query that clearly asks for information based on the assigned task. \\
Guidelines: The query can be answered by the content provided in the following given paragraph. The query should reflect a realistic information need in the disaster management domain. Avoid generic or overly broad questions—make the query specific and grounded in actual scenarios (e.g., logistics, policies, actions). Use language inspired by real-world usage, such as what a policymaker, journalist, or emergency planner might ask. Output only a single JSON object with a \texttt{user\_query} field. No extra formatting, documents, or explanations. Be clear, informative, and realistic!"\\
\hline
\end{tabular*}
\caption{Prompt templates for the user query generation for QAdoc-related search task. The \texttt{clarity} placeholder takes values: clear, understandable with some effort, and ambiguous. The \texttt{difficulty} placeholder includes: elementary school, high school, college, and PhD. For \texttt{query\_length}, possible values are: less than 10 words, 5 to 20 words, less than 20 words, at least 50 words, and at least 150 words. The \texttt{num\_words} placeholder takes values such as: at least 100 words, at least 200 words, at most 50 words, and 50 to 150 words.}
\label{tab:QAdoc_query_prompt}
\end{table*}

\begin{table*}[ht]
\centering
\small
\renewcommand{\arraystretch}{1.2}
\setlength{\tabcolsep}{4pt}
\begin{tabular*}{\textwidth}{@{\extracolsep{\fill}}p{\textwidth}@{}}
\hline
\textbf{Information Need Generation Stage}\\
\hline
"Brainstorm a list of entity retrieval tasks where the goal is to retrieve tweets that mention or provide relevant information about one or more entities (e.g., organizations, people, locations, events) found in the query. Here are a few examples to inspire your thinking: Given a query referencing “UNICEF,” retrieve tweets about their emergency relief efforts. Given a query about “Cyclone Mocha,” retrieve tweets reporting its impact or aftermath. Given a query that mentions “World Health Organization,” retrieve tweets that discuss their role in disaster health responses. Given a query with “Manila,” retrieve tweets about disaster conditions or relief actions in that location. Guidelines: Each task should be a single sentence describing a situation where an entity is referenced and tweets related to that entity should be retrieved. Focus on disaster management-related entities such as emergency response agencies, international organizations, locations, events, or key figures. Encourage diversity in topics: ground response, aid distribution, weather events, infrastructure failure, etc. Your output should be a JSON list of about 3 strings, each describing a different NER Twitter retrieval task. No explanations or extra formatting. Be concise, diverse, and realistic."\\
\hline
\textbf{User Query Generation Stage}\\
\hline
"You have been assigned an entity-tweet retrieval task: \verb|{task}|. Your mission is to write one example for this task in JSON format. The JSON object must include: \\
\;\;- \texttt{query}: a sentence that mentions one or more disaster management-related entities. \\
\;\;- \texttt{positive\_tweet}: a tweet that provides relevant and informative content about the mentioned entity or entities. \\
Guidelines: The query should clearly mention a recognizable entity tied to the disaster domain. The positive\_tweet should be informal, observational, or emotional—realistic Twitter-style language providing relevant details about the entity. All content should be inspired by disaster management-related themes such as rescue missions, weather events, humanitarian aid, response coordination, etc. Your output must be a single JSON object only. No explanation, no formatting beyond JSON. Keep it realistic and natural in tone."\\
\hline
\end{tabular*}
\caption{Prompt templates for the user query generation for Twitter-related search task. The \texttt{clarity} placeholder takes values: clear, understandable with some effort, and ambiguous. The \texttt{difficulty} placeholder includes: elementary school, high school, college, and PhD. For \texttt{query\_length}, possible values are: less than 10 words, 5 to 20 words, less than 20 words, at least 50 words, and at least 150 words. The \texttt{num\_words} placeholder takes values such as: at least 100 words, at least 200 words, at most 50 words, and 50 to 150 words.}
\label{tab:Twitter_query_prompt}
\end{table*}

\begin{table*}[ht]
\centering
\small
\renewcommand{\arraystretch}{1.2}
\setlength{\tabcolsep}{4pt}
\begin{tabular*}{\textwidth}{@{\extracolsep{\fill}}p{\textwidth}@{}}
\hline
\textbf{Information Need Generation Stage}\\
\hline
"Brainstorm a list of fact-checking retrieval tasks where the goal is: Given a claim, retrieve documents that either support or refute the claim, while distinguishing them from topically similar documents that do not address the claim's veracity. Here are a few examples to guide your ideas: Given a claim about the effectiveness of early warning systems during floods, retrieve documents that either support or refute the claim. Given a claim about the number of people displaced by a recent earthquake, retrieve evidence that verifies or challenges it. Given a claim about the government's relief distribution timeline, retrieve text that affirms or contradicts the stated timeline. Given a claim about the relationship between climate change and disaster frequency, retrieve relevant supporting or refuting content. Guidelines: Each task should be one sentence and describe what the claim is about and what kind of evidence is needed (support or refute). Base the topics on real-world domains such as disaster management, humanitarian aid, policy, climate, health impacts, etc. The tasks should vary in specificity and format (e.g., statistical claim, causal claim, factual assertion). Your output should be a JSON list of about 3 strings, each string representing a distinct fact-checking retrieval task. Output only the list. No explanations. Be creative and diverse in topic!"\\
\hline
\textbf{User Query Generation Stage}\\
\hline
"You have been assigned a fact-checking retrieval task: \verb|{task}|. Your mission is to write one fact-checking retrieval instance in JSON format. The JSON object must contain: \\
\;\;- \texttt{claim}: a short, factual or semi-factual statement (assertion) related to the task. \\
\;\;- \texttt{positive\_document}: a paragraph that supports or refutes the claim. \\
Guidelines: The claim should be \verb|{query_length}|, \verb|{clarity}|. The positive document must clearly support or refute the claim—either is acceptable. The claim should be clear, concise, and specific—not overly vague or too broad. Use examples from realistic disaster management-related content: climate events, emergency response, humanitarian relief, damage estimates, etc. All positive documents should be \verb|{num_words}| long. All generated content should be in English no matter the provided content language is. Both the claim and documents must be understandable with \verb|{difficulty}| level education. Output only a single JSON object. No additional text. Be precise and creative!"\\
\hline
\end{tabular*}
\caption{Prompt templates for the user query generation for fact-checking related search task. The \texttt{clarity} placeholder takes the values: clear, understandable with some effort, and ambiguous. The \texttt{difficulty} placeholder includes: elementary school, high school, college, and PhD. The \texttt{query\_length} placeholder accepts values such as: less than 10 words, 5 to 20 words, at least 10 words, at least 20 words, and at least 50 words. The \texttt{num\_words} placeholder includes: at most 15 words, at most 50 words, 50 to 150 words, at most 100 words, and at least 100 words.}
\label{tab:factcheck_prompt_query}
\end{table*}

\begin{table*}[ht]
\centering
\small
\renewcommand{\arraystretch}{1.2}
\setlength{\tabcolsep}{4pt}
\begin{tabular*}{\textwidth}{@{\extracolsep{\fill}}p{\textwidth}@{}}
\hline
\textbf{Information Need Generation Stage}\\
\hline
"Brainstorm a list of Natural Language Inference (NLI) retrieval tasks. In these tasks, the objective is: Given a premise sentence from a paragraph (e.g., about disaster management), retrieve a hypothesis sentence that is logically entailed by the premise. Here are a few examples to inspire your creativity: Given a sentence describing a government emergency response, retrieve a hypothesis that reflects an outcome or implication of that action. Given a statement about climate-induced hazards, retrieve a hypothesis summarizing the likely impact. Given a factual description of infrastructure damage, retrieve a hypothesis about the services affected. Given a claim about disaster preparedness strategies, retrieve a hypothesis that is logically supported by it. Guidelines: Each task description should be one sentence and should clearly specify the type of premise and the nature of the entailed hypothesis. Tasks should be generalizable across topics but inspired by domains such as climate, crisis response, risk, logistics, etc. Be diverse in topic and formality: from news-like to academic to conversational. Your output should be a JSON list of about 3 strings, each describing a different NLI retrieval task. Output only the list of task descriptions, no explanations."\\
\hline
\textbf{User Query Generation Stage}\\
\hline
"You have been assigned an NLI retrieval task: \verb|{task}|. Your mission is to write one example for this task in JSON format. The JSON object must include: \\
\;\;- \texttt{premise}: a sentence drawn or inspired from a paragraph (e.g., about disaster management). \\
\;\;- \texttt{entailed\_hypothesis}: a sentence that must logically follow from the premise. \\
Guidelines: The premise should be \verb|{query_length}|, \verb|{clarity}|. Use realistic examples from domains like climate risk, emergency response, infrastructure, health, or logistics, etc. Ensure the entailed hypothesis is non-trivial and clearly follows from the premise. Avoid word-for-word overlap between the sentences unless necessary for clarity. \texttt{entailed\_hypothesis} should be \verb|{num_words}| long. All generated content should be in English no matter the provided content language is. All contents require \verb|{difficulty}| level education to understand and should be diverse in terms of topic and length. Output a single JSON object only. Do not explain yourself or add anything else. Be creative and accurate!"\\
\hline
\end{tabular*}
\caption{Prompt templates for the user query generation for NLI-related search task. The \texttt{clarity} placeholder takes values: clear, understandable with some effort, and ambiguous. The \texttt{difficulty} placeholder includes: elementary school, high school, college, and PhD. The \texttt{query\_length} placeholder accepts values such as: less than 10 words, 5 to 20 words, at least 20 words, at least 50 words, and at least 150 words. The \texttt{num\_words} placeholder includes: less than 10 words, 5 to 20 words, at least 20 words, at least 50 words, and at most 50 words.}
\label{tab:nli_prompt_query}
\end{table*}

\begin{table*}[t]
\centering
\small
\renewcommand{\arraystretch}{1.2}
\setlength{\tabcolsep}{4pt}
\begin{tabular*}{\textwidth}{@{\extracolsep{\fill}}p{\textwidth}@{}}
\hline
\textbf{Information Need Generation Stage}\\
\hline
"Brainstorm a list of similar sentence retrieval tasks where the goal is: Given a sentence, retrieve other sentences that express the same or very similar meaning (paraphrases or semantically equivalent expressions). Here are a few examples to inspire your ideas: Given a sentence describing the impact of a flood, retrieve other sentences that paraphrase or closely restate the same impact. Given a sentence about the steps taken during emergency evacuation, retrieve sentences that express the same process using different wording. Given a sentence about climate-related disasters increasing in frequency, retrieve other sentences conveying the same trend. Given a factual statement about relief distribution, retrieve sentences that express the same fact using alternate phrasing. Guidelines: Each task should be written in one sentence and describe the kind of sentence (source) and what type of similar sentences should be retrieved. Focus on disaster management-related themes such as risk, policy, action, climate, or aid—but vary topics for diversity. Your output should be a JSON list of about 3 strings, each one describing a distinct similar sentence retrieval task. Output the list only. No explanations. Be creative and precise."\\
\hline
\textbf{User Query Generation Stage}\\
\hline
"You have been assigned a similar sentence retrieval task: \verb|{task}|. Your mission is to write one example for this task in JSON format. The JSON object must contain: \\
\;\;- \texttt{query}: a single sentence that expresses a specific idea. \\
\;\;- \texttt{positive}: a sentence that expresses the same meaning as the query (semantic equivalence or high similarity). \\
Guidelines: The query should be \verb|{query_length}|, \verb|{clarity}|. The query and positive should be semantically equivalent, possibly using different wording or structure. Avoid copy-paste or trivial rewordings—be realistic and diverse. Use examples inspired by real-world disaster management-related content: emergency protocols, environmental impact, infrastructure, humanitarian response, etc. All positive documents should be \verb|{num_words}| long. All generated content should be in English no matter the provided content language is. Both the query and documents must be understandable with \verb|{difficulty}| level education. Output only a single JSON object. No explanation. Make it high-quality and realistic!"\\
\hline
\end{tabular*}
\caption{Prompt templates for the user query generation for STS-related search task. The \texttt{clarity} placeholder takes values: clear, understandable with some effort, and ambiguous. The \texttt{difficulty} placeholder includes: elementary school, high school, college, and PhD. The \texttt{query\_length} placeholder accepts values such as: less than 10 words, 5 to 20 words, at least 50 words, and at most 50 words. The \texttt{num\_words} placeholder includes: less than 10 words, 5 to 20 words, at least 50 words, and at most 50 words.}
\label{tab:STS_prompt_query}
\end{table*}

\begin{table*}[ht]
\centering
\small
\renewcommand{\arraystretch}{1.2}
\setlength{\tabcolsep}{4pt}

\caption{LLM relevance judgment prompt templates for QA, QAdoc, and Twitter-related search tasks.}
\label{tab:QA_eva_prompt_template}
\end{table*}

\begin{table*}[ht]
\centering
\small
\renewcommand{\arraystretch}{1.2}
\setlength{\tabcolsep}{4pt}
\begin{tabular*}{\textwidth}{@{\extracolsep{\fill}}p{\textwidth}@{}}
\hline
\textbf{[1] Zero‐shot Direct Scoring}\\
\hline
\emph{System prompt:} You are a search quality rater evaluating evidence for fact-checking and outputting JSON.\\ [1ex]
\emph{User prompt:} Given a claim and a passage, score relevance ($0$–$3$): $0$ = Irrelevant,
$1$ = Related (no help),
$2$ = Relevant (unclear/mixed),
$3$ = Direct support/refutation.\\[1ex]
\emph{Important:} $1$ if related but no help,
$2$ if important info + noise,
$3$ if clearly supports/refutes.\\[1ex]
\textbf{Claim:} \texttt{\{Query\}},
\textbf{Passage:} \texttt{\{Passage\}}. Consider intent, support/refutation ($M$), trustworthiness ($T$), then decide final score ($O$). Output MUST be JSON: \texttt{\{"final\_score": <0\_to\_3>\}}\\
\hline

\textbf{[2] Chain‐of‐Thought Decomposed Reasoning}\\
\hline
\textbf{PHASE 1 – Answer Presence Prediction:}\\
Instruction: Given a passage and a claim, predict whether the passage includes information that supports or refutes the claim by producing either “Yes” or “No”.\\
\hline
\end{tabular*}
\caption{LLM relevance judgment prompt templates for FC-related search tasks.}
\label{tab:factcheck_prompt_templates}
\end{table*}

\begin{table*}[ht]
\centering
\small
\renewcommand{\arraystretch}{1.2}
\setlength{\tabcolsep}{4pt}
\begin{tabular*}{\textwidth}{@{\extracolsep{\fill}}p{\textwidth}@{}}
\hline
\textbf{[1] Zero‐shot Direct Scoring}\\
\hline
\emph{System prompt:} You are a rater evaluating entailment. Output only the final score in JSON.\\[1ex]
\emph{User prompt:} Given a premise and hypothesis, score entailment ($0$–$3$): $0$ = Not entailed/Contradicted,
$1$ = Related not entailed,
$2$ = Mostly entailed,
$3$ = Perfectly entailed.\\[1ex]
\emph{Important:} $1$ if related but not inferable,
$2$ if captures important implied content but not fully,
$3$ if clearly/fully supported.\\[1ex]
\textbf{Premise:} \texttt{\{Query\}},
\textbf{Hypothesis:} \texttt{\{Passage\}}. Consider premise implications, logical following ($E$), info gaps, then decide final score ($O$). Output MUST be JSON: \texttt{\{"final\_score": <0\_to\_3>\}}\\
\hline

\textbf{[2] Chain‐of‐Thought Decomposed Reasoning}\\
\hline
\textbf{PHASE 1 – Answer Presence Prediction:}\\
Instruction: Given a hypothesis and a premise, predict whether the hypothesis is entailed by the premise by producing either “Yes” or “No”.\\
\hline
\end{tabular*}
\caption{LLM relevance judgment prompt templates for NLI-related search tasks.}
\label{tab:nli_prompt_templates}
\end{table*}

\begin{table*}[ht]
\centering
\small
\renewcommand{\arraystretch}{1.2}
\setlength{\tabcolsep}{4pt}
\begin{tabular*}{\textwidth}{@{\extracolsep{\fill}}p{\textwidth}@{}}
\hline
\textbf{[1] Zero‐shot Direct Scoring for STS}\\
\hline
\emph{System prompt:} You are a semantic‐similarity rater. Output JSON with a score ($0$–$5$) where $0$ = unrelated, $5$ = semantically equivalent.\\[0.5ex]
\emph{User prompt:} Rate the semantic similarity $0$–$5$: $0$ = unrelated \(\mid\)
$1$ = slight \(\mid\)
$2$ = partial \(\mid\)
$3$ = moderate \(\mid\)
$4$ = high \(\mid\)
$5$ = equivalent\\[0.5ex]
\textbf{Sentence A:} \texttt{\{input1\}} \textbf{Sentence B:} \texttt{\{input2\}}\\[0.5ex]
Output JSON: \texttt{\{"final\_similarity\_score": <0\_to\_5>\}}\\
\hline
\end{tabular*}
\caption{LLM relevance judgment prompt templates for STS-related search tasks.}
\label{tab:sts_prompt_template}
\end{table*}

\begin{table*}[!htbp]
\centering
\small
\renewcommand{\arraystretch}{1.1}
\setlength{\tabcolsep}{4pt}
\begin{tabular*}{\textwidth}{@{\extracolsep{\fill}}l lrrrrrr@{}}
\hline
\textbf{Search Intent} & \textbf{Event Type} & \textbf{rel=0} & \textbf{rel=1} & \textbf{rel=2} & \textbf{rel=3} & \textbf{rel=4} & \textbf{rel=5} \\
\hline
\multirow{8}{*}{FC}
  & Bio   & 17987 & 3034 & 1952 & 1014 & 0 & 0 \\
  & Chem  & 16345 & 3845 & 2746 & 1656 & 0 & 0 \\
  & Env   & 16739 & 4243 & 3235 & 1588 & 0 & 0 \\
  & Extra & 16867 & 3748 & 2436 & 1312 & 0 & 0 \\
  & Geo   & 20151 & 4126 & 2549 & 1038 & 0 & 0 \\
  & MH    & 18671 & 4283 & 2601 & 1115 & 0 & 0 \\
  & Soc   & 20757 & 3287 & 2707 & 1099 & 0 & 0 \\
\hline
\end{tabular*}
\caption{Distribution of qrels scores \textup{rel=0} through \textup{rel=5} for each search task in DisastIR. ``rel'' represents relevance score. Only STS-related search task is labeled in 6 levels, with others labeled in 4 levels.}
\label{tab:score_distribution}
\end{table*}

\begin{table*}[ht]
\centering
\small
\renewcommand{\arraystretch}{1.1}
\setlength{\tabcolsep}{4pt}
\begin{tabular*}{\textwidth}{@{\extracolsep{\fill}}l lrrrrrr@{}}
\hline
\textbf{Search Intent} & \textbf{Event Type} & \textbf{rel=0} & \textbf{rel=1} & \textbf{rel=2} & \textbf{rel=3} & \textbf{rel=4} & \textbf{rel=5} \\
\hline
\multirow{8}{*}{FC}
  & Bio  &   1 &  11 &  19 & 169 &   0 &   0 \\
  & Chem &   0 &   5 &  15 & 180 &   0 &   0 \\
  & Env  &   0 &   2 &  14 & 184 &   0 &   0 \\
  & Extra&   1 &   6 &  24 & 169 &   0 &   0 \\
  & Geo  &   1 &   9 &  21 & 169 &   0 &   0 \\
  & MH   &   0 &   4 &  26 & 170 &   0 &   0 \\
  & Soc  &   1 &   2 &  17 & 180 &   0 &   0 \\
\hline
\end{tabular*}
\caption{Distribution of query-generated relevant document relevance scores \textup{rel-0} through \textup{rel-5}}
\label{tab:positive_doc_score_dist}
\end{table*}

\begin{table*}[ht]
  \centering
  \small
  \renewcommand{\arraystretch}{1.15}
  \setlength{\tabcolsep}{3pt}
  \begin{tabular*}{\textwidth}{@{\extracolsep{\fill}}@{\hspace{5pt}}llll r@{\hspace{5pt}}}
    \hline
    \textbf{Name}       & \textbf{Intent}    & \textbf{Link}                                                      & \textbf{\#} \\
    \hline
    MS Macro            & QA                 & \url{https://huggingface.co/datasets/microsoft/ms_marco}           & 400                    \\
    TriviaQA            & QA                 & \url{https://huggingface.co/datasets/sentence-transformers/trivia-qa-triplet} & 400      \\
    ALLNLI              & NLI                & \url{https://huggingface.co/datasets/sentence-transformers/all-nli}          & 400      \\
    XNLI                & NLI                & \url{https://huggingface.co/datasets/mteb/xnli/viewer/en}           & 400                    \\
    STSB                & STS                & \url{https://huggingface.co/datasets/sentence-transformers/stsb}              & 400      \\
    ClimateFever        & FC         & \url{https://huggingface.co/datasets/tdiggelm/climate_fever}       & 400                    \\
    \hline
  \end{tabular*}
  \caption{Overview of selected open-source datasets in LVHL. ``\#'' represents the number of selected queries in the corresponding dataset.}
  \label{tab:dataset-overview}
\end{table*}

\begin{table*}[!tp]
\centering
\small
\renewcommand{\arraystretch}{1.1}
\begin{tabular*}{\textwidth}{@{\extracolsep{\fill}}@{\hspace{5pt}}l l l l c c c@{\hspace{5pt}}}
\hline

\textbf{Model}    & \textbf{Param}   & \textbf{Size}  & \textbf{Base}   & \textbf{Embed.} & \textbf{MTEB} & \textbf{Arch.} \\

\textbf{Name}     & \textbf{Size}    & \textbf{Bin}   & \textbf{Model}  & \textbf{Size}   & \textbf{Rank} &               \\
\hline
inf-retriever-v1                       & 7B    & XL     & gte-Qwen2-7B-instruct               & 3584 & 1  & decoder \\
NV-Embed-v2                           & 7B    & XL     & Mistral-7B-v0.1                     & 4096 & 2  & decoder \\
inf-retriever-v1-1.5b                  & 1.5B  & XL     & gte-Qwen2-1.5B-instruct             & 1536 & 3  & decoder \\
Linq-Embed-Mistral          & 7B    & XL     & E5-mistral-7b-instruct              & 4096 & 4  & decoder \\
NV-Embed-v1                           & 7B    & XL     & Mistral-7B-v0.1                     & 4096 & 5  & decoder \\
SFR-Embedding-Mistral             & 7B    & XL     & E5-mistral-7b-instruct              & 4096 & 6  & decoder \\
\hline
\end{tabular*}
\caption{Information of all evaluated models. ``--'' means no publicly available information is available.}
\label{tab:model_specs}
\end{table*}

\begin{table*}[ht]
\centering
\small
\renewcommand{\arraystretch}{1.1}
\begin{tabular*}{\textwidth}{@{\extracolsep{\fill}}@{\hspace{5pt}}l p{0.45\textwidth} r@{\hspace{5pt}}}
\hline
\textbf{Model Name} & \textbf{Link} & \textbf{License} \\
\hline
inf-retriever-v1                       & \url{https://huggingface.co/infly/inf-retriever-v1} & apache-2.0 \\
NV-Embed-v2                            & \url{https://huggingface.co/nvidia/NV-Embed-v2} & cc-by-nc-4.0 \\
inf-retriever-v1-1.5b                  & \url{https://huggingface.co/infly/inf-retriever-v1-1.5b} & apache-2.0 \\
Linq-Embed-Mistral                     & \url{https://huggingface.co/Linq-AI-Research/Linq-Embed-Mistral} & cc-by-nc-4.0 \\
NV-Embed-v1                            & \url{https://huggingface.co/nvidia/NV-Embed-v1} & cc-by-nc-4.0 \\
SFR-Embedding-Mistral                  & \url{https://huggingface.co/Salesforce/SFR-Embedding-Mistral} & cc-by-nc-4.0 \\
snowflake-arctic-embed-l               & \url{https://huggingface.co/Snowflake/snowflake-arctic-embed-l} & apache-2.0 \\
\hline
\end{tabular*}
\caption{HuggingFace model links and licenses for all evaluated models.}
\label{tab:model_links}
\end{table*}

\begin{table*}[t]
\centering
\small
\renewcommand{\arraystretch}{1.1}
\begin{tabular*}{\textwidth}{@{\extracolsep{\fill}}@{\hspace{5pt}}lccccccc@{\hspace{5pt}}}
\hline
 & QA & QAdoc & Twitter & FC & NLI & STS & Avg. \\
\hline
\multicolumn{8}{l}{\textbf{Alibaba-NLP-gte-Qwen2-1.5B-instruct}}\\
\hline
Biological                  & 70.58 & 60.91 & 68.01 & 61.43 & 55.36 & 74.59 & 65.15 \\ 
Chemical                    & 70.02 & 58.58 & 64.21 & 64.40 & 55.48 & 73.43 & 64.35 \\ 
Environmental               & 69.88 & 56.90 & 65.41 & 64.73 & 59.01 & 73.74 & 64.94 \\ 
Extraterrestrial            & 67.56 & 56.18 & 61.33 & 62.44 & 53.14 & 74.72 & 62.56 \\ 
Geohazard                   & 70.30 & 58.34 & 63.81 & 59.60 & 54.98 & 71.91 & 63.16 \\ 
Meteorological\&hydrological & 70.45 & 59.65 & 66.28 & 59.56 & 54.95 & 72.93 & 63.97 \\
\hline
\end{tabular*}
\caption{Performance of the first six evaluated models under six search intents and eight event types under the exact search setting. Part I}
\label{tab:baseline_models_group1}
\end{table*}

\begin{table*}[t]
\centering
\small
\renewcommand{\arraystretch}{1.1}
\begin{tabular*}{\textwidth}{@{\extracolsep{\fill}}@{\hspace{5pt}}lccccccc@{\hspace{5pt}}}
\hline
 & QA & QAdoc & Twitter & FC & NLI & STS & Avg. \\
\hline
\multicolumn{8}{l}{\textbf{BAAI-bge-small-en-v1.5}}\\
\hline
Biological                  & 58.08 & 52.41 & 28.12 & 55.48 & 35.55 & 66.01 & 49.28 \\ 
Chemical                    & 56.12 & 48.37 & 27.78 & 57.54 & 32.94 & 64.88 & 47.94 \\ 
Environmental               & 60.67 & 51.59 & 24.29 & 56.89 & 37.56 & 64.23 & 49.21 \\ 
Extraterrestrial            & 56.59 & 51.14 & 23.53 & 56.96 & 32.45 & 65.57 & 47.71 \\ 
Geohazard                   & 54.58 & 48.79 & 22.84 & 53.02 & 30.27 & 63.40 & 45.48 \\ 
Meteorological\&hydrological & 56.82 & 51.60 & 24.47 & 54.81 & 32.21 & 62.84 & 47.13 \\
\hline
\end{tabular*}
\caption{Performance of evaluated models under six search intents and eight event types under the exact search setting. Part II}
\label{tab:baseline_models_group2}
\end{table*}

\begin{table*}[t]
\centering
\small
\renewcommand{\arraystretch}{1.1}
\begin{tabular*}{\textwidth}{@{\extracolsep{\fill}}@{\hspace{5pt}}lccccccc@{\hspace{5pt}}}
\hline
 & QA & QAdoc & Twitter & FC & NLI & STS & Avg. \\
\hline
\multicolumn{8}{l}{\textbf{Snowflake-snowflake-arctic-embed-m-v1.5}}\\
\hline
Biological                  & 20.43 & 30.04 & 17.22 & 48.66 & 43.01 & 63.04 & 37.07 \\ 
Chemical                    & 23.35 & 29.02 & 19.92 & 48.41 & 41.47 & 64.66 & 37.80 \\ 
Environmental               & 28.52 & 33.59 & 21.38 & 51.31 & 48.85 & 65.02 & 41.45 \\ 
Extraterrestrial            & 25.42 & 24.03 & 17.19 & 46.50 & 41.84 & 65.34 & 36.72 \\ 
Geohazard                   & 25.97 & 30.09 & 16.87 & 46.30 & 41.72 & 62.68 & 37.27 \\ 
Meteorological\&hydrological & 22.59 & 28.84 & 17.24 & 48.22 & 42.64 & 63.29 & 37.14 \\
\hline
\end{tabular*}
\caption{Performance of evaluated models under six search intents and eight event types under the exact search setting. Part III}
\label{tab:baseline_models_group3}
\end{table*}

\begin{table*}[t]
\centering
\small
\renewcommand{\arraystretch}{1.1}
\begin{tabular*}{\textwidth}{@{\extracolsep{\fill}}@{\hspace{5pt}}lccccccc@{\hspace{5pt}}}
\hline
 & QA & QAdoc & Twitter & FC & NLI & STS & Avg. \\
\hline
\multicolumn{8}{l}{\textbf{intfloat-e5-base-v2}}\\
\hline
Biological                  & 67.77 & 62.64 & 57.97 & 62.07 & 46.84 & 73.89 & 61.86 \\ 
Chemical                    & 66.40 & 59.51 & 57.92 & 63.12 & 45.14 & 73.75 & 60.98 \\ 
Environmental               & 67.34 & 62.93 & 59.94 & 64.35 & 49.39 & 74.33 & 63.05 \\ 
Extraterrestrial            & 63.25 & 63.36 & 55.28 & 63.11 & 44.20 & 74.21 & 60.57 \\ 
Geohazard                   & 65.05 & 60.12 & 55.61 & 60.40 & 43.29 & 73.40 & 59.64 \\ 
Meteorological\&hydrological & 65.83 & 62.57 & 58.66 & 62.12 & 44.30 & 72.80 & 61.04 \\
\hline
\end{tabular*}
\caption{Performance of evaluated models under six search intents and eight event types under the exact search setting. Part IV}
\label{tab:baseline_models_group4}
\end{table*}

\begin{table*}[t]
\centering
\small
\renewcommand{\arraystretch}{1.1}
\begin{tabular*}{\textwidth}{@{\extracolsep{\fill}}@{\hspace{5pt}}lccccccc@{\hspace{5pt}}}
\hline
 & QA & QAdoc & Twitter & FC & NLI & STS & Avg. \\
\hline
\multicolumn{8}{l}{\textbf{snowflake-arctic-embed-m-long}}\\\hline
Biological                  & 19.40 & 10.57 & 21.85 & 39.39 & 41.02 & 52.71 & 30.82 \\ 
Chemical                    & 19.97 & 10.58 & 21.06 & 36.59 & 40.40 & 55.39 & 30.66 \\ 
Environmental               & 21.57 & 11.81 & 21.01 & 34.99 & 48.55 & 55.06 & 32.16 \\ 
Extraterrestrial            & 21.22 & 11.89 & 17.32 & 35.89 & 41.66 & 56.66 & 30.77 \\ 
Geohazard                   & 21.78 & 8.92  & 17.99 & 35.32 & 40.09 & 52.50 & 29.44 \\ 
Meteorological\&hydrological & 19.71 & 10.04 & 18.89 & 36.12 & 40.67 & 54.56 & 30.00 \\
\hline
\end{tabular*}
\caption{Performance of evaluated models under six search intents and eight event types under the exact search setting. Part V}
\label{tab:baseline_models_group5}
\end{table*}

\end{document}